\documentclass[12pt,oneside, a4paper]{article}

\pdfoutput=1  
 
\ifx\pdfoutput\undefined
\usepackage[dvips,bookmarks=false]{hyperref}	
\else
\usepackage{hyperref}	
\fi
\usepackage{orcidlink}
\usepackage{xcolor}
\definecolor{linkcolor}{rgb}{0.0, 0.47, 0.75}
\definecolor{citecolor}{rgb}{1.0, 0.5, 0.0}
\hypersetup{
  linkcolor  = linkcolor,
  citecolor  = linkcolor,
  urlcolor   = linkcolor,
  colorlinks = true
}

\usepackage{graphicx}
\usepackage{subfigure}
\usepackage{amssymb}
\usepackage{cite}
\usepackage{bm}
\usepackage{amsmath,amsthm}
\usepackage{cleveref}
\usepackage{siunitx}
\usepackage{slashed}
\numberwithin{equation}{section}

\newcommand{\capdef}{}
\newcommand{\mycaption}[2][\capdef]{\renewcommand{\capdef}{#2}%
       \caption[#1]{{\footnotesize #2}}}

\begin{document}

\begin{titlepage}

\begin{center}

\vspace*{2cm}
        {\large\bf
Oscillating Neutrinos vs. Oscillating Scalars:\\
Constraining Scalar Dark Matter-Induced
Neutrino Mass}
        
\vspace{1cm}

\renewcommand{\thefootnote}{\fnsymbol{footnote}}

{\bf Kierthika Chathirathas~\orcidlink{0009-0005-5041-9867}},\footnote[1]{kierthika.chathirathas@kit.edu}
{\bf Sabya Sachi Chatterjee~\orcidlink{0000-0003-1730-2788}},\footnote[2]{sabya.chatterjee@kit.edu}
{\bf Thomas Schwetz~\orcidlink{0000-0001-7091-1764}}\footnote[3]{schwetz@kit.edu}
\vspace{5mm}

{\it%
  {Institut f\"ur Astroteilchenphysik, Karlsruher Institut f\"ur Technologie (KIT),\\
    Hermann-von-Helmholtz-Platz 1, 76344 Eggenstein-Leopoldshafen, Germany}}

\today
  
\vspace{8mm} 

\abstract{We consider the hypothesis that neutrino masses are generated by a coupling to an ultra-light (pseudo-)scalar field, which provides the dark matter in the universe. This leads to time-varying neutrino masses with a frequency set by the dark matter mass, with implications for neutrino oscillation data. We use that dark matter is in a virialised state in the galaxy and provide a detailed discussion of the relevant time scales. Using data from the T2K, RENO and JUNO experiments, we show that for dark matter masses smaller than about $3\times 10^{-8}$~eV down to the smallest allowed dark matter mass of about $10^{-21}$~eV only a fraction of between 9\% to 54\% of the total neutrino mass can arise from the coupling to the background scalar, depending on the value of the scalar mass. Future data from JUNO may improve these limits down to 1\% in certain regions of scalar masses. We focus on a real scalar field, but most of our results hold also for a complex scalar.
}

\end{center}

\end{titlepage}

\renewcommand{\thefootnote}{\arabic{footnote}}
\setcounter{footnote}{0}

\tableofcontents

\section{Introduction}

The origin of neutrino mass and the nature of dark matter remain outstanding open questions in particle physics. An intriguing possibility is that these two issues are connected. In the following we will consider the hypothesis that neutrino masses emerge from a coupling to an ultra-light scalar dark matter field. Such a dark matter candidate is well-motivated, in particular in the context of pseudo-Goldstone dark matter and axion-like particles. In the following we refer to this broad class of dark matter simply as ``scalar dark matter''. The impact of a neutrino coupling to scalar dark matter for neutrino oscillations has been investigated by a number of authors \cite{Berlin:2016woy,Krnjaic:2017zlz,Brdar:2017kbt,Liao:2018byh,Capozzi:2018bps,Choi:2019zxy,Ge:2018uhz,Dev:2020kgz,Losada:2021bxx,Losada:2022uvr,Losada:2023zap,Gherghetta:2023myo,Cheek:2025kks,Delgadillo:2025wxw}, see also \cite{Dev:2022bae,Huang:2022wmz,ChoeJo:2023ffp,Ge:2024ftz,FTAiroldi:2026hmu} for related phenomenology. In Ref.~\cite{Sen:2023uga}, Smirnov and Sen proposed a concrete model and identified two possible regions in parameter space, corresponding to scalar masses $m_\phi \simeq 10^{-10}-10^{-9}$~eV and coupling constants $g\simeq 10^{-10}-10^{-9}$ or $m_\phi \simeq 10^{-21}$~eV and $g\simeq 10^{-20}$, where neutrino mass could be explained entirely by the coupling to the background dark matter field. 

In our work we perform a model-independent analysis, considering dark matter masses $m_\phi \lesssim 10^{-8}$~eV. In this mass range, the coupling of neutrinos to the background field leads to time-varying neutrino masses on time scales relevant for neutrino oscillation experiments. This will lead to distortions of the oscillation probability as a function of neutrino energy and can therefore be constrained by data. We will consider a selective set of oscillation experiments and show that their observations are incompatible with the hypothesis that neutrino masses emerge entirely from a coupling to scalar dark matter. We are going to introduce a phenomenological parameterization, interpolating smoothly between standard vacuum masses and fully scalar dark matter-induced neutrino masses and set upper bounds on the dark matter contribution. 

In order to predict the effect of the neutrino--dark scalar coupling on oscillation phenomenology, one has to specify the state of the scalar field in the galaxy. We will adopt the random phase model, see e.g., \cite{Widrow:1993qq,Centers:2019dyn,Hui:2020hbq}, to describe the scalar field corresponding to a virialised dark matter system. This allows for a careful consideration of the coherence properties of the field, assuming a typical dark matter velocity distribution. We provide a detailed discussion of the emerging time scales, which are set by $1/m_\phi$, $1/(\bar v m_\phi)$ and $1/(\bar v^2 m_\phi)$, where $m_\phi$ is the scalar mass and $\bar v \sim 220$~km/s is the typical dark matter velocity at our location in the Milky Way. These time scales have to be compared with the relevant experimental time scales $T_{\rm exp}$ (total exposure time of an experiment) and $T_\nu$ (the time a neutrino needs to travel from the source to the detector). Depending on the relations of the time scales, we will identify different regimes, where scalar field-induced time variations can either be neglected (because they are too slow) or are fully averaged out (for fast oscillating terms). 

In this work we focus on a limited set of oscillation experiments, namely JUNO, RENO and T2K, where for the latter we use only the disappearance channel. These experiments allow to constrain all oscillation parameters except the complex phase using only disappearance data. Moreover, they are dominated by vacuum oscillations, which we use as a simplifying assumption in our analysis. 

\begin{figure}[t]
  \centering
  \includegraphics[width=\textwidth]{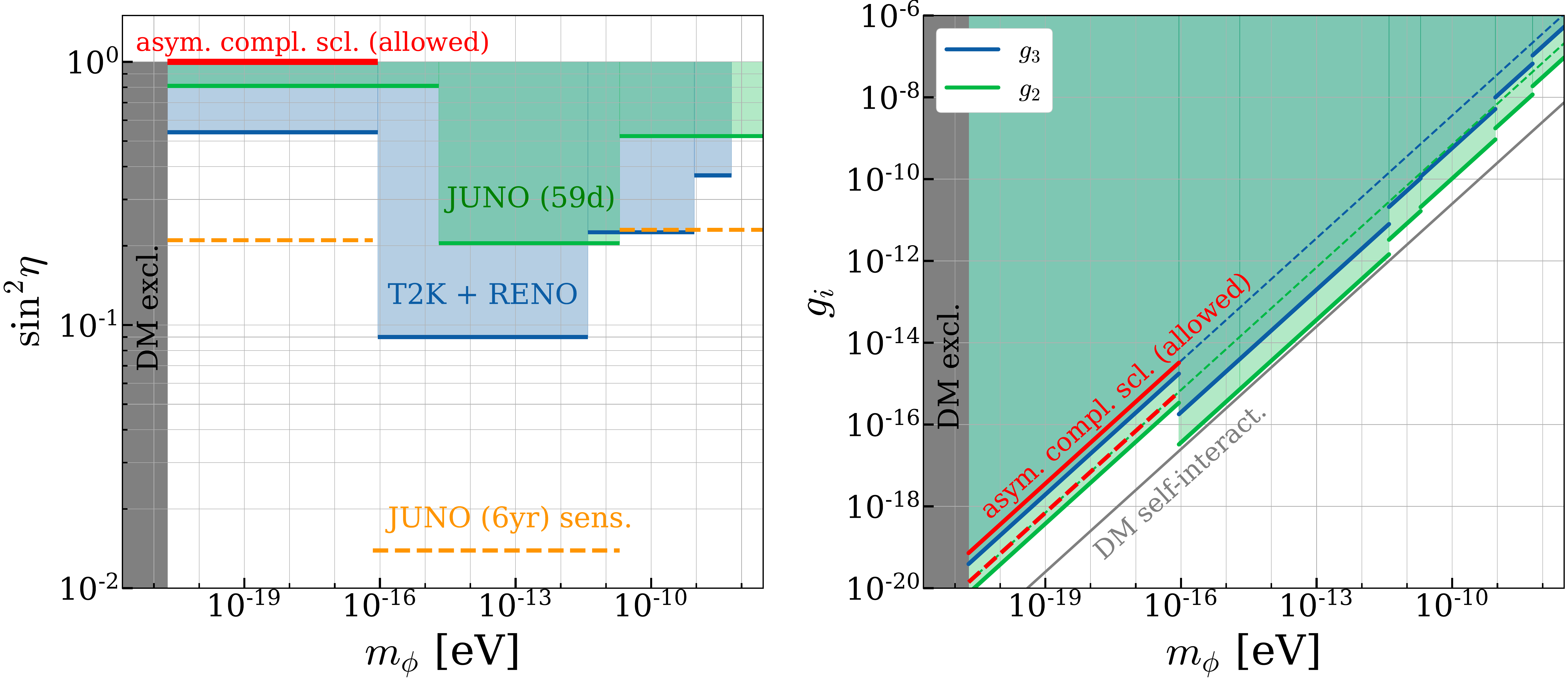}
  \mycaption{\textit{Left:} Upper limits on the fraction of scalar dark matter-induced neutrino mass as a function of the scalar mass $m_\phi$. Blue and green shaded regions are excluded  at $3\sigma$ by T2K+RENO and JUNO, respectively. The dashed orange lines indicate the sensitivity of 6~years of JUNO data. \textit{Right:} Upper limits on the scalar--neutrino coupling constants $g_2$ (green) and $g_3$ (blue) as a function of the scalar mass, see \cref{eq:Lphi}, assuming for the lightest neutrino mass $m_1 = 0$. Thin-dashed lines indicate the values of the couplings needed to provide 100\% of neutrino masses. The light-grey solid line shows the upper limit from cosmological structure formation due to loop-induced dark matter self-interactions \cite{Aghaie:2026bqq,Cembranos:2018ulm}. The red bars in both panels indicate the allowed parameter space in case of a fully asymmetric complex scalar field. The grey-shaded region in both panels are excluded by the lower bound on the dark matter mass \cite{Zimmermann:2024xvd}. }
  \label{fig:summary}
\end{figure}

As an executive summary, we provide the main results of our analysis in \cref{fig:summary}. The left panel shows the constraints on the relative contribution of the scalar-induced neutrino mass, parametrized by the parameter $\sin^2\eta$. As visible from the figure, we can exclude a 100\% dark origin of neutrino mass over many decades in scalar mass $m_\phi$. In the right panel, we show the corresponding constraints on the coupling constants $g_{2,3}$ between the scalar and neutrinos, contributing to the neutrino mass states $m_2$ and $m_3$, respectively, assuming that the lightest neutrino mass $m_1$ is zero. 

Our results apply for a real scalar or pseudo-scalar field. Most of the shaded parameter space in the figure is also excluded for a complex scalar field, with the only exception being the regions indicated in red, which are allowed for a fully asymmetric complex scalar background (consisting \textit{only} of particles \textit{or} anti-particles) and providing 100\% of neutrino masses. The discontinuities in the limits in \cref{fig:summary} are due to our averaging approximations in the different time scale regimes and the bounds should not be taken at face value close to the jumps. 

In the right panel of \cref{fig:summary} we show also a bound on the scalar--neutrino coupling derived in \cite{Aghaie:2026bqq} by considering the effective scalar potential induced by loop effects due to the scalar--neutrino interaction. This leads to dark matter self-interactions which are strongly constrained by cosmological observations \cite{Cembranos:2018ulm}. A comparable but slightly weaker bound has been derived by similar arguments in \cite{Dev:2022bae} from Milky Way satellites. Our bounds from neutrino oscillation data are comparable in strength to these loop-induced cosmological bounds and very complementary in nature. 

We note that the scalar dark matter-induced neutrino mass scales with the dark matter density as $\sqrt{\rho_{\rm DM}}$. This would lead to a very large mass of neutrinos in the early Universe, e.g., at recombination or BBN, if the effective description remains valid in that regime. Therefore, one needs to invoke a mechanism to suppress the scalar--induced neutrino mass throughout large part of the cosmological history, see e.g., \cite{Huang:2022wmz,Sen:2023uga,Sen:2024pgb}.

The outline of the remainder of the paper is as follows. In \cref{sec:framework} we introduce the framework, discuss the scalar-field properties in the galaxy and identify three regimes, depending on the relation between scalar-field and experimental time scales. Furthermore, we discuss the time-averaging effect on vacuum neutrino oscillation probabilities, depending on the regime. In \cref{sec:constraints} we present the results of our numerical analysis of the T2K and RENO data (\cref{sec:T2K-RENO}) and of the first data release of JUNO (\cref{sec:JUNO}), and we provide a sensitivity estimate for the final 6 year exposure of JUNO (\cref{sec:JUNOsens}). We summarize our results and provide further discussions in \cref{sec:discussion}. In the main part of the text we focus on a real scalar field, while a complex scalar field is discussed in appendix~\ref{app:complex}. Technical details on the T2K analysis are provided in appendix~\ref{app:t2k}.

\section{Framework}\label{sec:framework}

We consider a classical, non-relativistic scalar field $\phi$ providing the dark matter in the universe. Considering the local dark matter density in the galaxy of $\rho_{\rm DM} \simeq 0.4$~GeV/cm$^3$, and typical dark matter velocities of $v\simeq 200$~km/s, we find occupation numbers of
\begin{equation}
  \frac{\rho_{\rm DM}}{m_\phi} \lambda_{\rm dB}^3 \simeq \mathcal{O}(1)
  \left(\frac{40\,{\rm eV}}{m_\phi}\right)^4 \,,
\end{equation}
with $\lambda_{\rm dB} = 2\pi/(m_\phi v)$ denoting the deBroglie wavelength. We will be interested in scalar fields with masses $m_\phi \lll 1$~eV, and therefore we expect huge occupation numbers in the galaxy, which justifies the treatment of $\phi$ as a classical field. In the non-relativistic limit, the amplitude of the field is related to the local energy density by
\begin{equation}
  \rho_{\rm DM} = \frac{1}{2}m_\phi^2\langle\phi^2\rangle \,.
\end{equation}
Therefore, we parametrize the scalar field as
\begin{equation}\label{eq:phiA}
  \phi(\vec x,t) = \frac{\sqrt{2\rho_{\rm DM}}}{m_\phi} A(\vec x,t) \,,
\end{equation}
with $\langle A(\vec x,t)^2\rangle = 1$. The space and time dependence of $A(\vec x,t)$ depends on the coherence properties of the field, which we will discuss below.

We assume the following effective interaction between neutrinos and a real scalar field $\phi$:
\begin{equation}\label{eq:Lphi}
  \mathcal{L}_\phi = \frac{1}{2} \sum_{i=1}^3 g_i \phi \overline{\nu^c_{iL}}\nu_{iL} + \text{h.c.}\,, 
\end{equation}
with $\nu_{iL}$ denoting the left-chiral neutrino field corresponding to the mass-eigenstate with mass $m_i$. 
Given the classical nature of the background field $\phi$, this acts precisely as a Majorana mass term for neutrinos, with $m_i^{\rm dark} = g_i\phi$, where we can chose $g_i$ real without loss of generality. This interaction is not gauge invariant and therefore is an effective operator. 
Here we remain agnostic about possible UV completions and consider \cref{eq:Lphi} as an effective interaction describing the relevant phenomenology. In \cref{eq:Lphi} we have chosen a Majorana-type coupling for the sake of definiteness. Our discussion applies equally in the case of a Dirac-like coupling replacing $\overline{\nu_{iL}^c}\nu_{iL} \to 2 \overline{\nu_{iR}}\nu_{iL}$. 

Note also that a coupling of the type \cref{eq:Lphi} covers both scalar and pseudo-scalar fields (including well-motivated pseudo-Goldstone boson axion-like particles).
In the main text we focus on a single real scalar field (taking $\phi$ and $g_i$ real). The case of a complex scalar is considered in appendix~\ref{app:complex}. As discussed there, most of our constraints will apply also in that case, with some exceptions, in particular for a highly asymmetric complex scalar background.

In our work we will ask the question whether the interaction in \cref{eq:Lphi} can be the sole origin of neutrino mass. In that limit \cref{eq:Lphi} defines the neutrino mass basis. To perform the phenomenological analysis it is convenient to have a continuous parameterization interpolating between standard vacuum neutrino masses and the scalar-field model. Therefore, we adopt the following ansatz for neutrino masses:
\begin{equation}\label{eq:m_nu}
  \mu_i(\vec x,t) = m_i^{\rm vac} + m_i^{\rm dark}(\vec x,t) =
  m_i^0\left[c_\eta^2 + s_\eta^2 A(\vec x,t)\right] \,,
\end{equation}
where $c_\eta \equiv \cos\eta$, $s_\eta \equiv \sin\eta$, and $0\le \eta \le \pi/2$ interpolates between pure vacuum mass and pure scalar field-induced mass. Here $m_i^0$ are constant mass parameters with 
 $m_i^{\rm vac} = m_i^0 c_\eta^2$ and 
\begin{equation}\label{eq:coupling_sqeta}
  \frac{g_i}{m_\phi} = \frac{m_i^0}{\sqrt{2\rho_{\rm DM}}}\sin^2\eta
  \,.
\end{equation}

An important assumption in our analysis is the proportionality of the couplings $g_i$ and the vacuum neutrino masses $m_i^{\rm vac}$. In a generic model we do not expect this particular relation between the two quantities. In general they can have different flavour structure and may not be diagonal in the same basis. We consider this ansatz as a phenomenological parameterization to interpolate between the two limiting cases of pure vacuum and pure scalar-induced neutrino masses, corresponding to $s^2_\eta = 0$ and 1, respectively. 

Under this assumption, the expression in the square bracket in \cref{eq:m_nu} containing all the space and time dependence becomes an overall factor of the neutrino mass matrix and PMNS mixing angles are constant in space and time. Neutrino oscillations will be only affected by a time and space dependence of the mass-squared differences:
\begin{align}\label{eq:dmu2}
  \Delta \mu^2_{ij}(\vec x, t) & \equiv \mu_i^2-\mu_j^2 =
  \Delta m^2_{ij} \left[c_\eta^2 + s_\eta^2 A(\vec x,t)\right]^2 \,,
\end{align}
where $\Delta m^2_{ij}  \equiv (m^0_i)^2-(m^0_j)^2$. This is the effect we are going to explore below.

Note that in configurations where the matter effect is important, our assumptions may not hold. The effective mass basis in matter will no longer coincide with the basis in which the scalar couplings $g_i$ are diagonal and effective mixing angles will become space and time dependent. This will lead to a rich phenomenology, which we leave for future investigation. Below we restrict our analysis to experimental configurations where the matter effect can be neglected.

\subsection{The scalar field in the galaxy}\label{sec:scalar-field}

In general, one expects a virialised system for the dark matter halo of our galaxy. In the context of scalar field dark matter, this can be described by the random phase model, e.g., \cite{Widrow:1993qq,Centers:2019dyn,Hui:2020hbq}, where the scalar field is a superposition of many momentum modes with random phases. 
In this case, $A(\vec x,t)$ from \cref{eq:phiA} is given by
\begin{equation}\label{eq:Axt}
  A(\vec x,t) = \sum_{\vec k}\alpha_{\vec k} \, a_{\vec k} \,\cos\left[ m_\phi t + \frac{\vec k^2t}{2m_\phi} - \vec k \vec x + \varphi_{\vec k} \right] \,.
\end{equation}
Here the coefficients $a_{\vec k}$ are proportional to the dark matter velocity distribution in the galaxy. Our default assumption is based on the standard halo model, and correspondingly we adopt a Maxwellian velocity distribution in the galactic rest frame, boosted into the earth laboratory frame:
\begin{align}\label{eq:SHM}
  a_{\vec k}^2 \propto \exp\left[ -\frac{(\vec v + \vec v_e)^2}{\bar v^2}\right] \,, \qquad
  \sum_{\vec k} a_{\vec k}^2 = 1 \,,
\end{align}
with $\vec k = m_\phi \vec v$ and $\vec v_e$ denotes the velocity of the earth with respect to the halo rest frame with $|\vec v_e| \approx \bar v = 220$~km/s. We neglect the cut-off introduced by the escape velocity as well as the motion of the earth around the sun. Furthermore, $\varphi_{\vec k}$ are random phases in $[0,2\pi]$ for each momentum mode $\vec k$, $\alpha_{\vec k}$ are Rayleigh-distributed random variables with the probability density $P[\alpha] = 2\alpha e^{-\alpha^2}$, and
\begin{align}
  \langle e^{i\varphi_{\vec k}}\rangle_{\rm ens} = 0 \,, \quad
  \langle e^{i(\varphi_{\vec k}-\varphi_{\vec k'})}\rangle_{\rm ens} = \delta_{\vec k \vec k'} \,, \quad
  \langle \alpha^2_{\vec k}\rangle_{\rm ens} = 1 \,.
\end{align}
The random numbers $\alpha_{\vec k}$ take into account that each momentum bin $\vec k$ contains many individual modes with random phases relative to each other, see e.g., \cite{Foster:2017hbq} for a derivation. We indicate here explicitly the ensemble average over different realisations of $\varphi_{\vec k}$ and $\alpha_{\vec k}$, to be distinguished from the time average considered below, although in general we expect the ergodic theorem to hold.

Note that in \cref{eq:Axt} we have adopted a real decomposition of the field, as appropriate for a real scalar field. The complex analogue is discussed in appendix~\ref{app:complex}, see \cref{eq:Axt-compl}.

A given neutrino oscillation experiment will be affected by the field value $\phi(\vec x, t)$ as a function of time during the run time of the experiment, at the position of the experiment $\vec x$.
We consider \cref{eq:Axt} in the rest frame of the earth, consistent with \cref{eq:SHM}. Therefore $\vec x$ describes the coordinates of the neutrinos propagating from the source to the detector. 
The term $\vec k\vec x$ can be neglected, as long as the deBroglie wave length of the scalar field is large compared to the source-detector distance $L = |\vec L|$, where $\vec L = \vec x_D - \vec x_S$, with $\vec x_{S,D}$ denoting the coordinates of neutrino source and detector:
\begin{align}\label{eq:dB}
\vec k\vec L = m_\phi \vec v \vec L \ll 2\pi \,.
\end{align}
Taking into account that $|\vec v| \sim \bar v \sim 10^{-3} c$, we notice the following time scales in \cref{eq:Axt}:
\begin{align}
  \tau_\phi &\equiv \frac{2\pi}{m_\phi} \approx 4 \,{\rm s}\,
  \left(\frac{10^{-15} \,{\rm eV}}{m_\phi}\right) \,, \label{eq:tau_phi}\\
  \tau_{\rm dB} &\equiv \frac{2\pi}{m_\phi \bar v} \simeq 10^3  \tau_\phi \,, \label{eq:tau_dB}\\
  \tau_v &\equiv \frac{2\pi}{m_\phi {\bar v}^2} \simeq 10^6  \tau_\phi \,. \label{eq:tau_v}
\end{align}
On the other hand, the experimental configuration of a given neutrino oscillation experiment determines also two time scales: $T_\nu$, the time a neutrino needs to travel from the source to the detector, i.e., $T_\nu = L$, and $T_{\rm exp}$, the total data taking time of the experiment, typically of order years. The values of $T_\nu$ and $T_{\rm exp}$ for the JUNO, RENO, and T2K experiments are summarized in \cref{tab:times}. For JUNO we consider the run time of 59.1~days from the first results published in \cite{JUNO:2025gmd} as well as the projected total run time of 6~years.

Depending on the relation between the scalar-field and the experimental times scales we divide the analysis in different regimes, in order of increasing scalar mass $m_\phi$:
\begin{equation}\label{eq:regimes}
\begin{split}    
\text{(R1):}\qquad  & \tau_v \gg T_{\rm exp} \gg \tau_\phi \gg T_\nu \,,\\
\text{(R2):}\qquad  & T_{\rm exp} \gg \tau_v \gg \tau_\phi \gg T_\nu \,,\\
\text{(R3):}\qquad  & T_{\rm exp} \gg \tau_v \gg \tau_{\rm dB} \gg T_\nu \gg \tau_\phi \,.  
\end{split}    
\end{equation}
The condition $\tau_{\rm dB} \gg T_\nu$ explicitly imposed in case 3 is equivalent to \cref{eq:dB} and ensures that the term $\vec k\vec x$  is just a constant for fixed $\vec k$, which can be absorbed in the random phase $\varphi_{\vec k}$. 

\begin{table}[t]
\centering
  \begin{tabular}{l@{\qquad}c@{\qquad}c}
    \hline\hline
    experiment & $T_{\rm exp}$ & $T_{\nu} $ \\
    \hline
    T2K  &  $\approx 1.6$ yr $\approx 5\times 10^{7}$ s 
         &  295 km$/c \approx 0.98 \times 10^{-3}$ s \\
    RENO &  3800 d $\approx 3.3\times 10^8$ s 
         &  $\approx 1.4$ km$/c \approx 4.7\times 10^{-6}$ s \\
    JUNO (59 d) &  59.1 d $\approx 5.1\times 10^6$ s 
      & 53 km$/c \approx 1.8\times 10^{-4}$ s \\
    JUNO (6 yr) &  6 yr $\approx 1.9\times 10^8$ s 
      & 53 km$/c \approx 1.8\times 10^{-4}$ s \\
    \hline\hline
  \end{tabular}  
  \mycaption{Experimental run time $T_{\rm exp}$ and $T_\nu = L/c$ for the three experiments JUNO, RENO, and T2K. For the latter, we quote an effective value of $T_{\rm exp}$ obtained by averaging over the run-time periods, weighted with the beam power and the $\bar\nu/\nu$ ratio \cite{T2K:2023smv}; see appendix~\ref{app:t2k} for details.}
  \label{tab:times}
\end{table}

\begin{figure}[t]
  \centering
  \includegraphics[width=\textwidth]{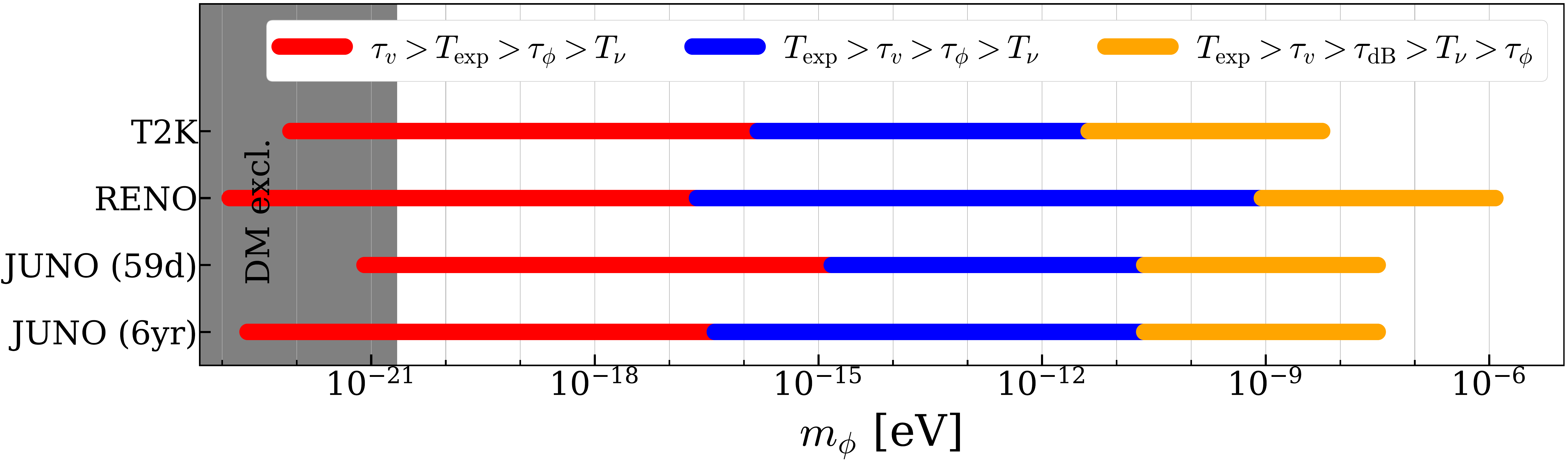}
  \mycaption{Illustration of the three regimes as a function of the scalar mass $m_\phi$ for the JUNO, RENO, and T2K experiments, using the time scales given in \cref{tab:times} and $\bar v = 220$~km/s. The grey shaded region is excluded by the model-independent lower bound on the dark matter mass from dwarf-galaxies \cite{Zimmermann:2024xvd}. Regime~1, 2, 3 is shown in red, blue, orange, respectively.}
  \label{fig:regimes}
\end{figure}

\Cref{fig:regimes} shows the three regimes for the JUNO, RENO, and T2K experiments as a function of the scalar mass. As indicated in the figure, the regime (R1) extends for all three experiments down to values of $m_\phi < 2\times 10^{-21}$, which is the model-independent lower bound on the dark matter mass from dwarf-galaxies \cite{Zimmermann:2024xvd}. Note that under additional assumptions on the dark matter production mechanism substantially stronger lower bounds on $m_\phi$ may apply, see, e.g.~\cite{Amin:2022nlh,Eberhardt:2025caq,Chathirathas:2025aan}. The important observation is that the regimes (R1), (R2), (R3) cover the huge range of dark matter masses from the model-independent lower bound up to masses of order $10^{-8}$~eV, for RENO even up to $10^{-6}$~eV. The dark matter lower bound $m_\phi > 2.2\times 10^{21}$~eV implies $\tau_\phi < 21$~days, and therefore we can always assume that $\tau_\phi < T_{\rm exp}$ and the scalar field oscillates many times during the run time of the experiments.

Under the assumption \cref{eq:dB}, the function in \cref{eq:Axt} becomes independent of $\vec x$ and is only a function of time: $A(\vec x,t)\to A(t)$. It is convenient to separate the fast and the slow time dependencies in the following way:
\begin{align}\label{eq:At}
  A(t) = \cos(m_\phi t) C(t) - \sin(m_\phi t) S(t) = R(t) \cos[m_\phi t + \gamma(t)] \,,
\end{align}
where
\begin{align}
  C(t) &= \sum_{\vec k}\alpha_{\vec k} \, a_{\vec k} \cos(\beta_{\vec k}) \,,\quad
  S(t) = \sum_{\vec k}\alpha_{\vec k} \, a_{\vec k} \sin(\beta_{\vec k}) \,, \label{eq:CS}\\
  R(t)^2 &= C(t)^2+ S(t)^2 \,,\quad \tan\gamma(t) = \frac{S(t)}{C(t)} \,, \quad 
  \beta_{\vec k}(t) = \frac{\vec k^2}{2m_\phi}t  + \varphi_{\vec k} \,.
  \label{eq:R_gamma_beta}
\end{align}
The quantities in \cref{eq:CS,eq:R_gamma_beta} are ``slow'' functions of time, with the typical frequency $2\pi/\tau_v \sim 10^{-6} m_\phi$, and they include the random properties of the scalar field encoded in $\alpha_{\vec k}$ and $\varphi_{\vec k}$.

\subsection{Neutrino oscillations} \label{sec:oscillations}

Let us now discuss the impact of time varying neutrino masses, \cref{eq:dmu2} for the neutrino oscillation probabilities. We first consider the regimes (R1) and (R2), where $\tau_\phi \gg T_\nu$, i.e., the neutrino mass does not change during the propagation from the source to the detector. However, during the run time of the experiment, each neutrino will ``see'' a different mass. In these cases, the oscillation probability will have the same form as for standard vacuum masses, but needs to be averaged over the data taking period of a given experiment. 

The time dependence will enter in the expression for the vacuum probability via terms
\begin{equation}
    \sin^2\frac{\Delta \mu^2_{ij}(t)L}{4E_\nu} = 
    \frac{1}{2}\left(1 - \cos\frac{\Delta \mu^2_{ij}(t)L}{2E_\nu}\right)  \,.
\end{equation}
We write
\begin{equation}\label{eq:cos_t}
    \cos\frac{\Delta \mu^2_{ij}(t)L}{2E_\nu} = \cos \Delta_{ij} [c_\eta^2 + s_\eta^2 A(t)]^2,
\end{equation}
where
\begin{equation}
    \Delta_{ij} \equiv \frac{\Delta m^2_{ij}L}{2E_\nu} \,,
\end{equation}
with the constant mass-squared difference $\Delta m^2_{ij}$ defined in \cref{eq:dmu2}. For $A(t)$ we adopt the expression in \cref{eq:At}. Let us discuss now the time average over 
\cref{eq:cos_t} in the two regimes (R1) and (R2).

In the \textbf{regime (R1)}, $\tau_v \gg T_{\rm exp}$ and therefore, $R$ and $\gamma$ take random but constant values over the whole run-time period of the experiment. Hence, in (R1) we cannot predict the specific value of $R$ and $\gamma$. They are determined by the random values of $\varphi_{\vec k}$ and $\alpha_{\vec k}$ and correspond to a particular coherence patch of the axion field in which the earth happens to be located during the run time of the experiment. $R$ follows a Rayleigh distribution with scale parameter $1/\sqrt{2}$, and $R^2$ is exponentially distributed with $\lambda = 1$. Hence, we have
\begin{align}\label{eq:Rmean}
    \langle R^2 \rangle_{\rm ens} = 1 \,,\quad \sigma(R^2) = 1 \,,\qquad
    \langle R \rangle_{\rm ens} = \frac{\sqrt{\pi}}{2}\approx 0.886 \,,\quad 
    \sigma(R) = \sqrt{1-\frac{\pi}{4}}\approx 0.463 \,,
\end{align}
with $\sigma(X)$ denoting the standard deviation of $X$. Given the sizeable standard deviations, the values can cover quite a large spread. The only relevant time dependence enters via the ``fast'' term $\cos(m_\phi t + \gamma)$. Averaging of these terms can be done with the help of the Jacobi-Anger identity \cite{Cheek:2025kks}. After dropping the fast-oscillating terms which average to zero we find
\begin{align}\label{eq:bessel}
   \left\langle \cos\frac{\Delta \mu^2_{ij}L}{2E_\nu} \right\rangle_{T_{\rm exp}} &=
    \cos\left[\Delta_{ij}\left(c_\eta^4 + \frac{s_\eta^4 R^2}{2}\right)\right] J_0(2c_\eta^2s_\eta^2 R \Delta_{ij})J_0\left(\frac{s_\eta^4R^2\Delta_{ij}}{2}\right) \nonumber\\
    &+2\text{Re}\left\{ \exp\left[i\Delta_{ij}\left(c_\eta^4 + \frac{s_\eta^4R^2}{2}\right)\right]
    \sum_{n=1}^\infty (-i)^n J_{2n}(2c_\eta^2s_\eta^2 R \Delta_{ij})J_n\left(\frac{s_\eta^4R^2\Delta_{ij}}{2}\right)
    \right\}
\end{align}
with $J_n$ denoting the Bessel functions of the first kind. For $c_\eta=0,s_\eta=1$, only the term $n=0$ remains and we recover the expression from \cite{Cheek:2025kks}; \cref{eq:bessel} generalizes their result to intermediate values of $s_\eta$. In practice, it suffices to take into account terms up to $n\le 5$.
Note that \cref{eq:bessel} does no longer depend on the phase $\gamma$, but it does depend on the fixed but unknown value of $R$.

In contrast, in \textbf{regime (R2)}, $\tau_v \ll T_{\rm exp}$, and therefore $R(t)$ and $\gamma(t)$ oscillate many times during the run time of the experiment and we need to average the oscillation probability also over their time dependence. We have verified by explicit numerical calculations that to good approximation this additional averaging can be implemented by replacing $R\to\langle R\rangle$ and $R^2\to\langle R\rangle^2$ in \cref{eq:bessel} with $\langle R\rangle = 0.886$  given in \cref{eq:Rmean}.\footnote{As a consequence of the ergodic theorem we have $\langle \cdot \rangle_{\rm ens} = \langle \cdot \rangle_{\rm time}$.} 

\begin{figure}[t]
  \centering
  \includegraphics[width=\textwidth]{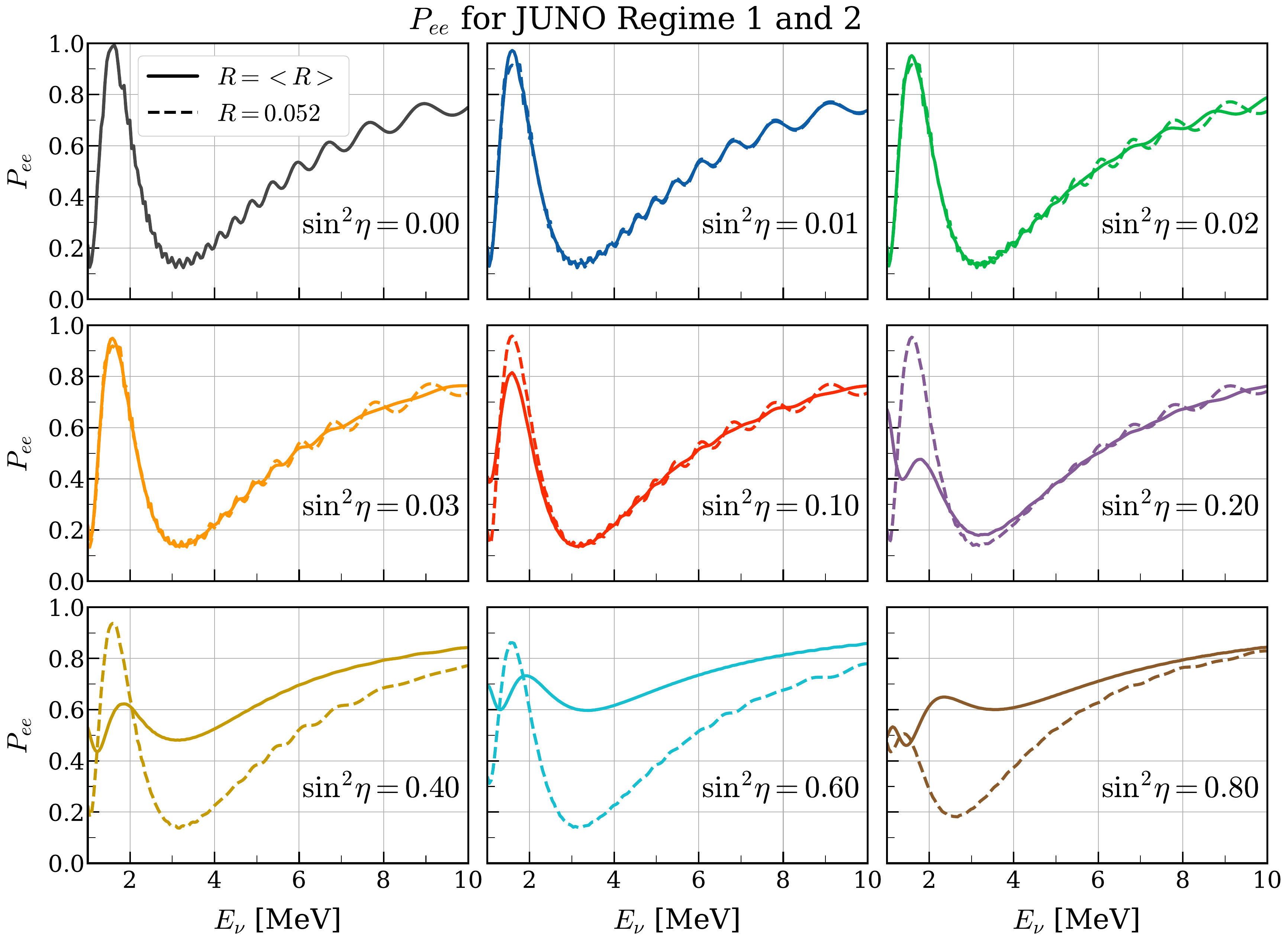}
  \mycaption{Oscillation probabilities for JUNO in regimes~1 and 2, see \cref{eq:regimes}, for $\sin^2\eta = 0, 0.01, 0.02, 0.03, 0.1, 0.2, 0.4, 0.6, 0.8$. Solid curves are for the mean value $R = \langle R\rangle = 0.886$, as relevant in regime~2. The dashed curves indicate the probabilities for $R = 0.052$ corresponding to the $3\sigma$ lower bound for $R$ in regime~1. Neutrino oscillation parameters for each value of $\sin^2\eta$ and each regime are optimized to best fit the data (see \cref{sec:constraints}).}
  \label{fig:probR12}
\end{figure}

In \cref{fig:probR12} we show the survival probability in JUNO as a function of neutrino energy for increasing values of the coupling to the scalar background, $\sin^2\eta$, with solid curves corresponding to (R2) and dashed curves apply to (R1), adopting the value $R = 0.052$, which is the $3\sigma$ lower bound derived from the Rayleigh distribution. We observe a strong distortion of the oscillation pattern if $\sin^2\eta \gtrsim 0.1$. The fast, $\Delta m^2_{31}$-induced oscillations get damped away already for $\sin^2\eta \gtrsim 0.1$. For larger values also the leading, $\Delta m^2_{21}$ driven oscillations get distorted. For $\sin^2\eta \gtrsim 0.4$, oscillation probabilities less than 0.4 cannot be obtained, which is strongly excluded by observations \cite{JUNO:2025gmd}. Note that for each value of $\sin^2\eta$ we use optimized oscillation parameters from the fit discussed below. For the $\nu_\mu$ survival probability relevant in T2K we find similar behaviour, with strong damping of oscillations for increasing coupling.

\bigskip

\begin{figure}[t]
  \centering
  \includegraphics[width=0.48\textwidth]{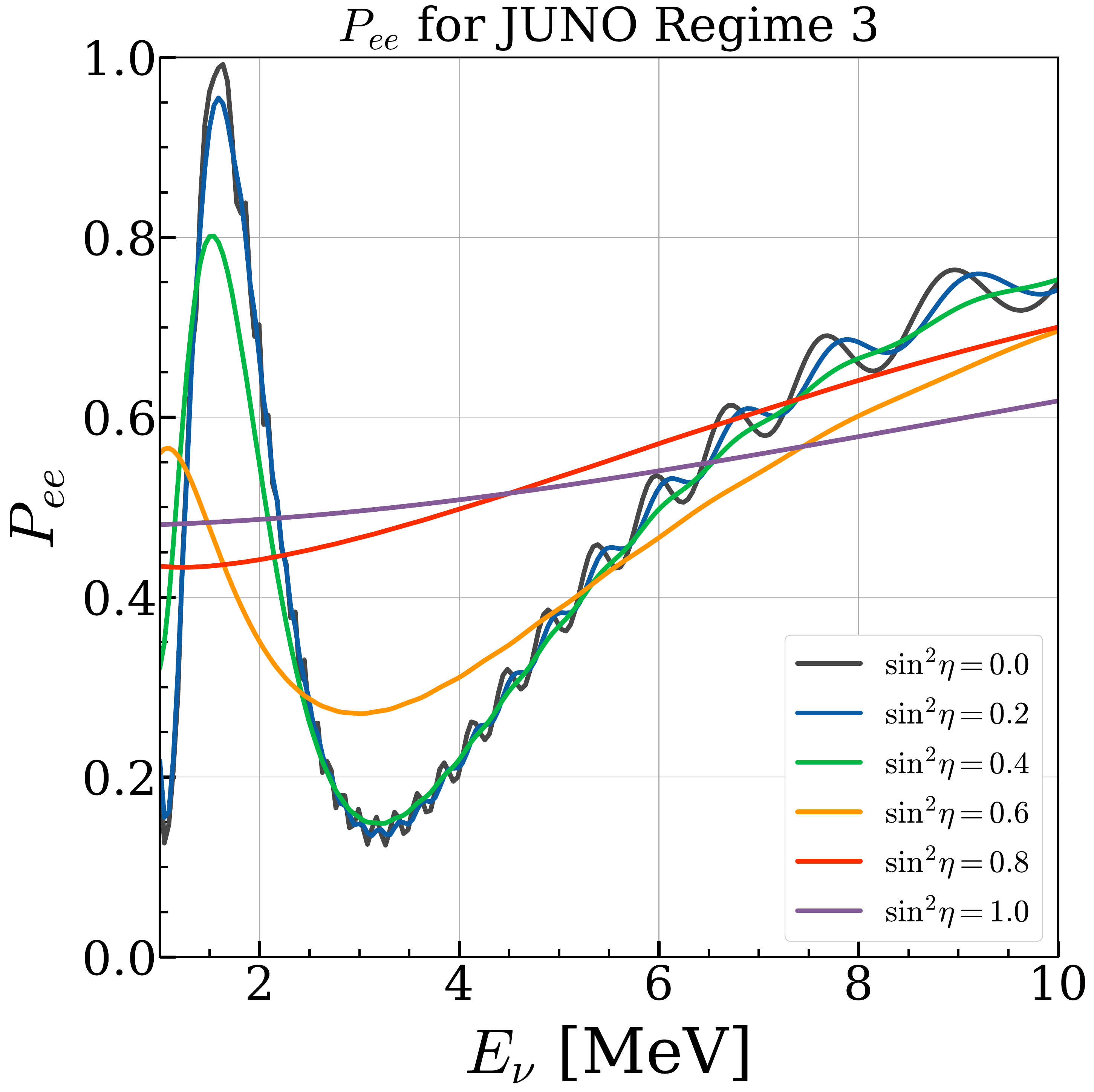}
  \includegraphics[width=0.48\textwidth]{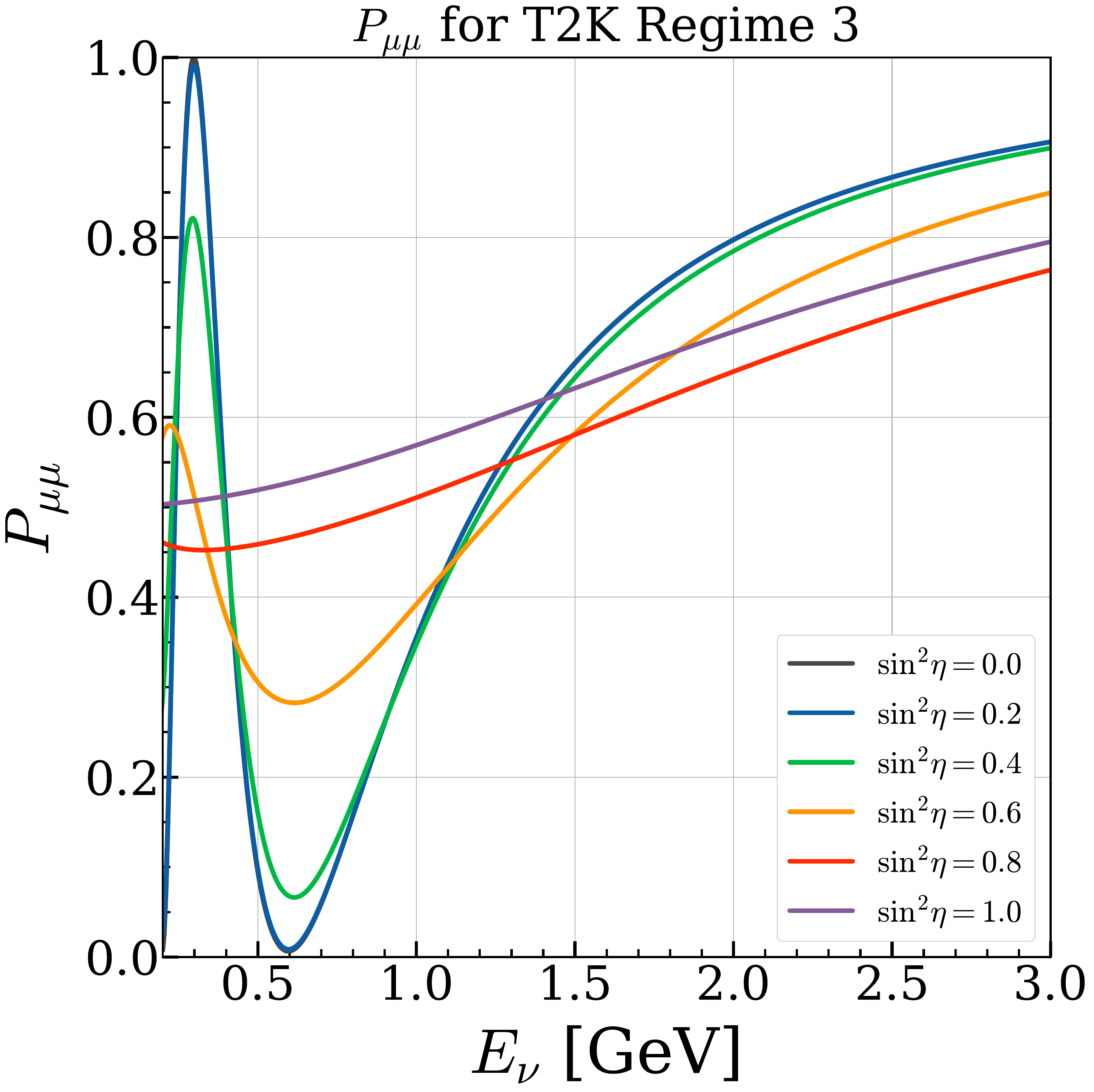}
  \mycaption{Probabilities for JUNO (left) and T2K (right) in regime~3 for different values of $\sin^2\eta$. Neutrino oscillation parameters for each value of $\sin^2\eta$ are optimized to best fit the data (see \cref{sec:constraints}).}
  \label{fig:probR3}
\end{figure}

Let us now discuss \textbf{regime (R3)}, where the neutrino mass oscillates with frequency $m_\phi$ many times during the propagation of each individual neutrino along the baseline of the experiment. Integrating the evolution equation of the mass eigenstates from the source to the detector leads then to an averaging of the oscillating terms in $\Delta \mu^2_{ij}$:
\begin{align}\label{eq:mu-R3}
    \int_t^{t+L} dt' \Delta \mu^2_{ij}(t') 
    \quad\to\quad 
    \Delta m^2_{ij} L \left(c_\eta^4 + \frac{s_\eta^4 R^2(t)}{2}\right) \,.
\end{align}
Since we have for (R3) that $T_{\rm exp}\gg \tau_v \gg T_\nu$, $R(t)$ will be constant for an individual neutrino along its propagation, however, will have a different value for each neutrino during the run time of the experiment. Hence, for (R3) the relevant averaging becomes
\begin{align}\label{eq:R3a}
   \left\langle \cos\frac{\Delta \mu^2_{ij}L}{2E_\nu} \right\rangle_{T_{\rm exp}} =
   \left\langle\cos\left[\Delta_{ij}\left(c_\eta^4 + \frac{s_\eta^4 R^2(t)}{2}\right)\right] \right\rangle_{T_{\rm exp}}
\end{align}
We can use that $R^2$ is exponentially distributed and the ergodic theorem to obtain
\begin{align}\label{eq:R3}
   \left\langle \cos\frac{\Delta \mu^2_{ij}L}{2E_\nu} \right\rangle_{T_{\rm exp}} =
   \frac{\cos(\Delta_{ij} c_\eta^4 + \alpha)}{\sqrt{1 + (\Delta_{ij} s_\eta^4/2)^2}}  \,,\qquad
   \tan\alpha = \frac{\Delta_{ij} s_\eta^4}{2} \,.
\end{align}
The resulting survival probabilities for the JUNO and T2K experiments are shown in \cref{fig:probR3}.  It is clear that also in this case, we get a strong distortion of the oscillatory pattern for large values of $\sin^2\eta$, clearly incompatible with observations. 

For \cref{eq:mu-R3} we used our assumption that vacuum masses $m_i^{\rm vac}$ and the couplings to the scalar $g_i$ are diagonal in the same basis, which defines the basis where the evolution Hamiltonian is diagonal (which requires also that Earth matter effects are negligible). In other cases, when off-diagonal couplings $g_{ij}$ are present, a parametric resonsance can occur when the scalar oscillation period $\tau_\phi$ matches the neutrino oscillation length, see \cite{Losada:2022uvr}. Under our assumptions stated above, this effect, however, does not occur and integrating the evolution equation only leads to the averaging indicated in \cref{eq:mu-R3}.

For even larger scalar masses, when $\tau_{\rm dB}$ or even $\tau_v$ becomes smaller than $T_\nu$, the averaging over $T_\nu=L$ in \cref{eq:mu-R3} would also affect $R(t)$ as well as additional space dependence via the term $\vec k\vec x$. Using $\langle R^2\rangle = 1$, then we can replace
\begin{align}\label{eq:mu-aver}
    \langle \Delta \mu^2_{ij}(t) \rangle_{T_\nu} 
    \quad\to\quad 
    \Delta m^2_{ij}  \left(c_\eta^4 + \frac{s_\eta^4}{2}\right) \,,
\end{align}
which becomes constant at the time scales $T_{\rm exp} \gg T_\nu$. Hence, within our framework there would be a complete degeneracy and vacuum oscillations cannot distinguish between vacuum and scalar-induced neutrino masses, as the effect can be compensated by readjusting $\Delta m^2_{ij}$. We note, however, that this holds only in our simplified scenario, where all neutrino masses are rescaled proportionally and mixing angles are unaffected. Furthermore, additional effects may appear in experiments for which the matter effect becomes relevant. Therefore, we limit our analysis to scalar masses $m_\phi \lesssim 10^{-8}$~eV (see \cref{fig:regimes}), such that the conditions for (R3) specified in \cref{eq:regimes} are fulfilled for the T2K and JUNO experiments.

In summary, \cref{eq:bessel,eq:R3} are the main results of this section and we will use them below for the phenomenological analysis. They predict a significant distortion of the oscillation pattern for a sizable coupling to the DM scalar field in all three regimes (R1), (R2), (R3).  

\subsection{Discussion of coherence properties}\label{sec:coherence}

For the analysis so far we have adopted the standard halo model for the dark matter distribution in the galaxy and the assumption of a virialised dark matter field, implemented by the random phase ansatz in \cref{eq:Axt}. The dark matter distribution is characterised by a typical velocity $\bar v$, which is about 220~km/s in the standard halo model. While this ansatz gives a realistic description of scalar dark matter, we briefly comment on the validity of our analysis in non-standard astrophysical scenarios, where the dark matter distribution may have different properties.

As long as the dark matter particles are virialised in the galaxy, the random phase ansatz \cref{eq:Axt} applies. For velocity distributions deviating from the standard halo model, the numerical value of the characteristic velocity $\bar v$ may be different from the generic value of 220~km/s. In this case, the transitions between the regimes (R1), (R2), (R3) will happen for different values of the scalar mass. However, the effect on neutrino oscillations will remain the same in each regime. In this sense, our analysis is robust and rather independent of the details of the dark matter distribution.

An exception is the case (R3) (with $\tau_\phi < T_\nu$) for a very cold dark matter distribution, such that $\tau_v > T_{\rm exp}$. This implies a velocity spread $\bar v < \sqrt{T_\nu/T_{\rm exp}} \sim 10^{-5}$ for JUNO and T2K and $\bar v \lesssim 10^{-7}$ for RENO. In other words, this corresponds to the situation where the scalar mass is large enough that the field oscillates fast while a single neutrino propagates, but at the same time the coherence time of the field is longer than the run time of the experiment. The extreme case of that scenario would be the limit $\bar v \to 0$ and $\tau_{\rm dB},\tau_v \to \infty$, i.e., a background scalar field consisting of just a single momentum mode, corresponding to a perfectly homogeneous scalar field oscillating coherently (Bose-Einstein condensate). In this case, $R$ will be always constant over the run time of the experiment. Our results for cases (R1) would apply also for the region denoted as (R2) above, extending up to the mass of the transition to (R3). However, for (R3), when $\tau_\phi < T_\nu$, the analysis would be different: the averaging in \cref{eq:R3} does not apply as $R$ is constant, and only the averaging over the fast $m_\phi$ oscillations in \cref{eq:mu-R3} applies. In that case a similar degeneracy as in \cref{eq:mu-aver} would appear and the same comments as stated there apply. We do not consider this special case further and leave it for future studies. 

\section{Constraints from neutrino oscillation data}
\label{sec:constraints}

In this section we present our numerical results using data from the T2K, RENO and JUNO experiments. This is a rather specific choice of data, motivated in the following way: One of our working assumptions is that the coupling to the scalar field is diagonal in the same basis as a possible vacuum neutrino mass matrix, and in particular it is diagonal in the propagation basis of neutrinos. If this requirement is fulfilled in vacuum, it will not hold if matter effects \cite{Mikheev:1986gs,Wolfenstein:1977ue} are important for neutrino propagation. Therefore, we restrict our analysis to experiments where matter effects can be neglected. This holds in very good approximation for the RENO reactor experiment. For T2K we only consider data from the $\nu_\mu$ and $\bar\nu_\mu$ disappearance channels. In contrast to the appearance channel, the disappearance survival probabilities are very accurately approximated by vacuum oscillations \cite{Denton:2024thm}, which justifies our framework neglecting matter effects. For the JUNO reactor experiment matter effects are small, at the percent level, but cannot be neglected to accurately fit the observed data already with present exposure \cite{Esteban:2026phq}. In our numerical analysis we do include standard matter effects to perform the fit, however, we assume that the scalar-induced effects do not modify the matter effect at leading order. 

Furthermore, this set of experimental data provides leading sensitivity to all three-flavour mixing parameters except the CP phase $\delta_{\rm CP}$
(see \cite{ParticleDataGroup:2026aaa} for the convention): $\Delta m^2_{31}$ and $\theta_{23}$ for T2K disappearance data, 
$\Delta m^2_{31}$ and $\theta_{13}$ for RENO, and
$\Delta m^2_{21}$ and $\theta_{12}$ for JUNO. We neglect the subleading sensitivity of T2K disappearance data to $\delta_{\rm CP}$ and for simplicity we restrict the analysis to the normal neutrino mass ordering. 
Hence, our choice of data represents a minimal set constraining the full PMNS mixing parameters (apart from $\delta_{\rm CP}$), while still respecting our requirement of vacuum propagation. In this sense our results are robust and conservative. It can be expected that in a more complete analysis including more data as well as matter effects, stronger bounds would emerge, which however, is beyond the scope of the present analysis. 

\subsection{T2K and RENO} \label{sec:T2K-RENO}

We start presenting our numerical results using data from the T2K and RENO experiments. Our analysis of RENO data is based on the final 3800~days exposure \cite{RENO:2024msr}. We are fitting the ratio of the events in the far and near detectors, with the appropriately weighted contribution of all reactors. Our analysis is well calibrated to accurately reproduce the official collaboration results \cite{RENO:2024msr} in the case of standard oscillations, based on \cite{nufit-6.1,Esteban:2018azc}. Our T2K analysis is based on disappearance data with $3.6\times 10^{21}$ POT exposure \cite{T2K:2023smv,T2K:2023mcm}. Details of our implementation are given in appendix~\ref{app:t2k}.

\begin{figure}[t]
  \centering
  \includegraphics[width=0.85\textwidth]{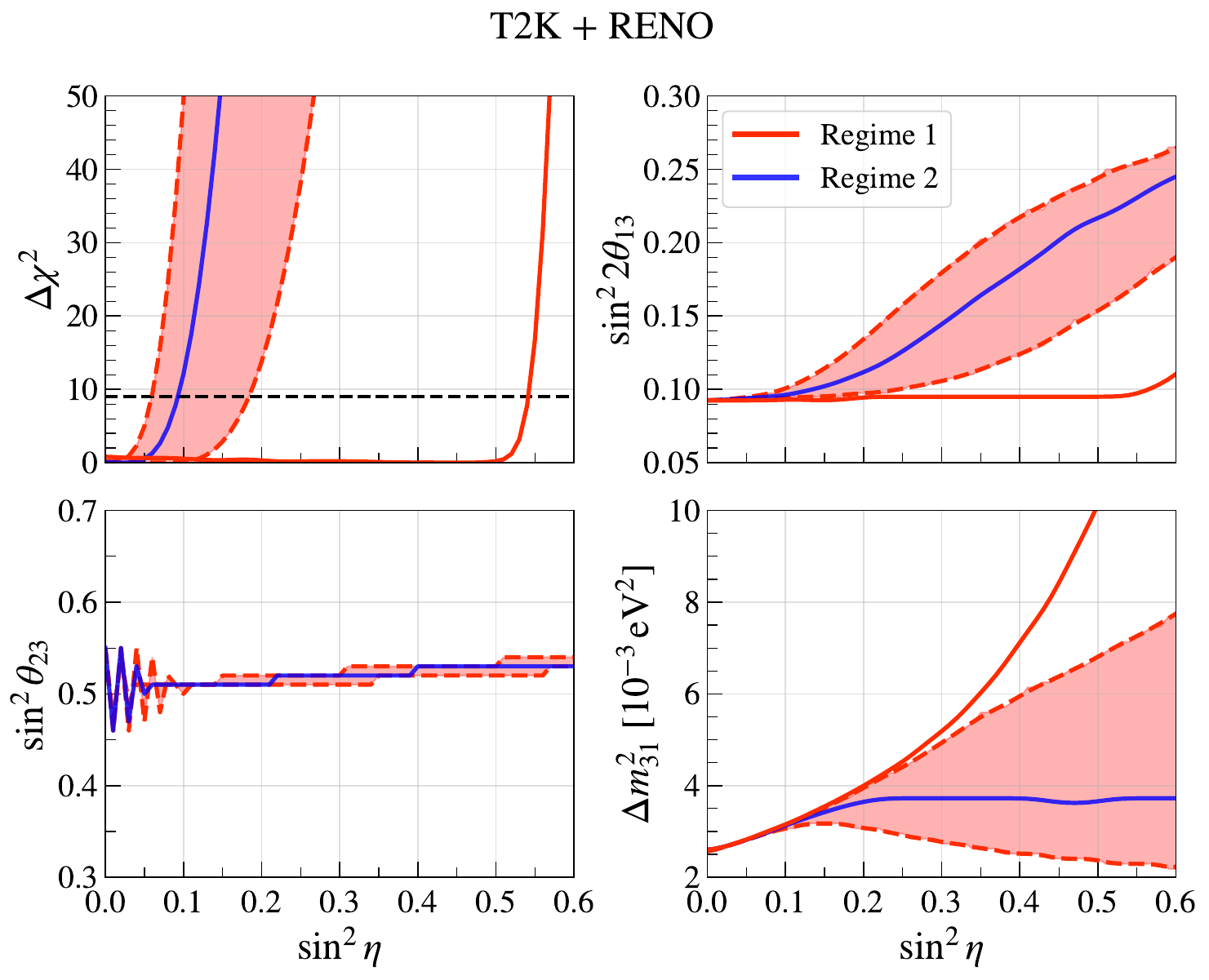}
  \mycaption{Combined analysis of T2K and RENO data for regimes 1 and 2. The upper left panel shows $\Delta\chi^2$ as a function of $\sin^2\eta$, the fraction of scalar-induced neutrino mass. The upper right and lower panels show the best-fit values of the oscillation parameters $\theta_{13}$, $\theta_{23}$, and $\Delta m_{31}^2$ for each value of $\sin^2\eta$. The blue curves in each plot corresponds to setting the $R$-parameter to $\langle R\rangle = 0.886$ as relevant for regime~2. Red curves refer to regime~1, where the dashed curves indicate the 
  $\pm \sigma(R)$ range for $R$, see \cref{eq:Rmean}, whereas the solid red curve corresponds to $R=0.052$, which is the one-sided lower bound at 99.73\%~CL as obtained from the Rayleigh distribution (not shown in the bottom-left panel).}
  \label{fig:T2K+RENO-R12}
\end{figure}

\Cref{fig:T2K+RENO-R12} shows our results for the combined T2K and RENO analyses assuming regimes~1 and 2 in terms of the fractional contribution of the scalar-induced neutrino mass parametrized by $\sin^2\eta$ as defined in \cref{eq:m_nu}. First we note that RENO on its own essentially cannot constrain $\sin^2\eta$. The reason is that in our conservative analysis we use only the far-to-near detector ratio. It turns out that in this ratio the averaging effects due to the background scalar oscillations nearly fully cancel, even for $\sin^2\eta \simeq 1$. Therefore, we show results only for the combined T2K and RENO data, with the main purpose of RENO being to provide a constraint on $\theta_{13}$, which is poorly constrained by T2K disappearance data. We have checked that otherwise the bound on $\sin^2\eta$ is fully dominated by T2K data.

The $\nu_\mu$ and $\nu_e$ disappearance channels relevant for T2K and RENO are consistently treated in a 3-flavour framework. For this analysis we fix the solar parameters $\theta_{12}$ and $\Delta m^2_{12}$ to their vacuum best-fit values. We have checked that varying their values according to the results of the JUNO analysis shown below as a function of $\sin^2\eta$ has a negligible impact on the T2K and RENO results. 

As discussed in \cref{sec:oscillations}, for regime~1 ($m_\phi \lesssim 10^{-16}$~eV) we cannot predict the value of the parameter $R$, which is determined from the random coherence patch of the background field the earth happens to be located in during the experiment. We show in the figure as shaded region between the red dashed curves results corresponding to the $\pm1\sigma$ range for $R$ according to \cref{eq:Rmean}. We see that small values of $R$ allow for larger fraction of scalar-induced mass. Therefore, in order to set a conservative upper limit on $\sin^2\eta$ in regime~1 at a given CL, we take the lower bound on $R$ at that CL, as obtained from the Rayleigh distribution with scale parameter $1/\sqrt{2}$, see discussion in \cref{sec:oscillations}. The corresponding $\chi^2$ curve is shown in the upper-left panel of \cref{fig:T2K+RENO-R12} for 99.73\%~CL as red solid curve. As expected from \cref{eq:bessel}, we see a significant weakening of the bound for such a low value of $R$, leading to $\sin^2\eta < 0.54$ at 99.73\%~CL.

In contrast, for regime~2 we have to average over many coherence patches during the run time of the experiment. We have checked that in this case it is a good approximation to set $R=\langle R \rangle = 0.886$. The results for regime~2 are shown as blue curves in \cref{fig:T2K+RENO-R12}, providing a significantly stronger limit of $\sin^2\eta < 0.09$ at 99.73\%~CL.

\begin{figure}[t]
  \centering
  \includegraphics[width=0.85\textwidth]{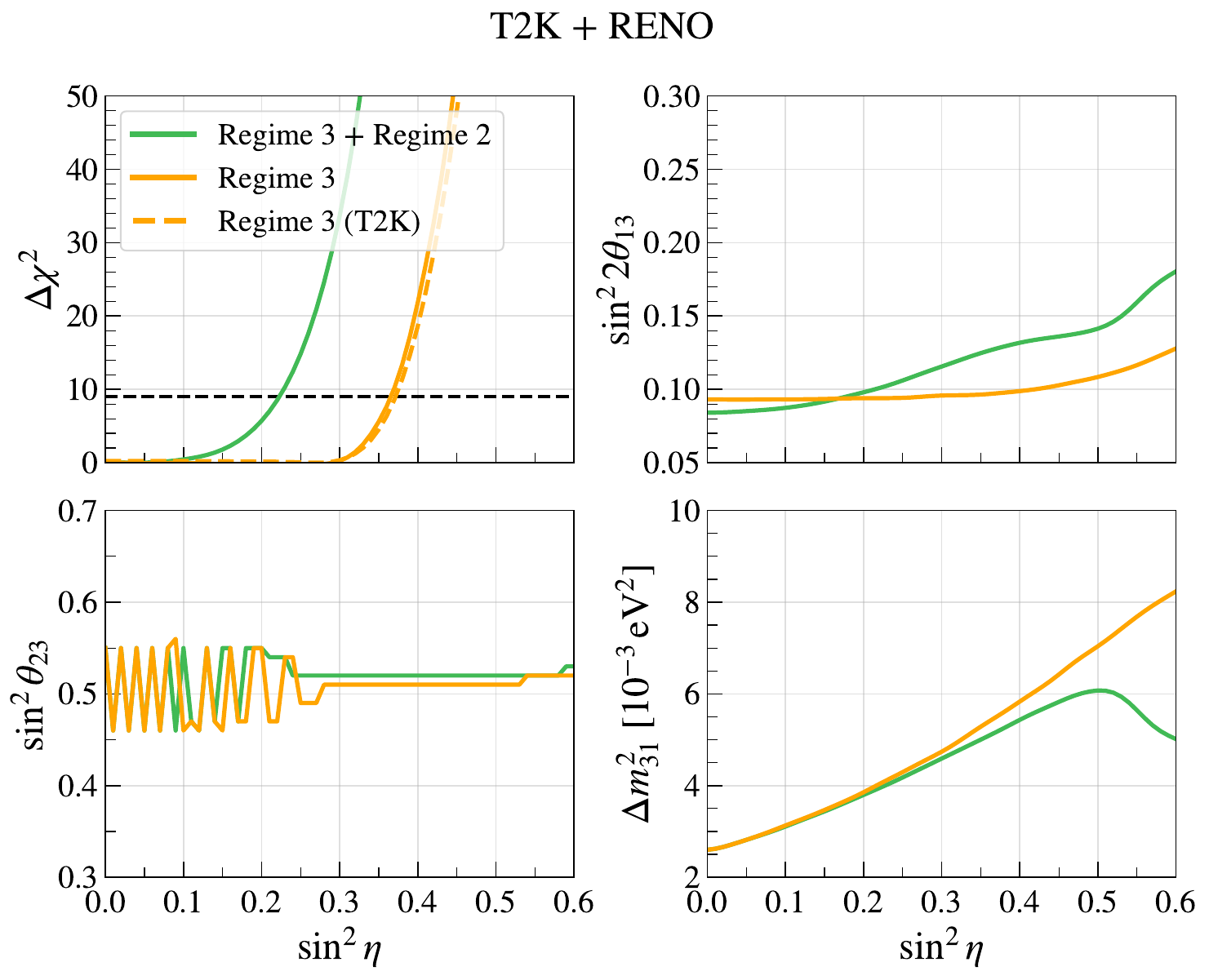}
  \mycaption{Combined analysis of T2K and RENO data for regime 3 (orange curves) and the mixed case, regime~3 for T2K and regime~2 for RENO (green curves). The upper-left panel shows $\Delta\chi^2$ as a function of $\sin^2\eta$; the orange-dashed curve corresponds to regime~3 for T2K only. The upper-right and lower panels show the best-fit values of the oscillation parameters $\theta_{13}$, $\theta_{23}$, and $\Delta m_{31}^2$ for each value of $\sin^2\eta$.}
  \label{fig:T2K+RENO-R3}
\end{figure}

\Cref{fig:T2K+RENO-R3} provides the results for regime~3 (orange curves), which applies in the mass range $10^{-9}\,{\rm eV} \lesssim m_\phi \lesssim 10^{-8}\,{\rm eV}$. As visible in \cref{fig:regimes}, in the range $3\times 10^{-12}\,{\rm eV} \lesssim m_\phi \lesssim 10^{-9}\,{\rm eV}$ T2K is in regime~3, whereas for RENO regime~2 applies. Therefore, we show in \cref{fig:T2K+RENO-R3} also this mixed case as green curve. In this case, there is actually a non-trivial synergy between the two experiments, and the combined limit (green curve) is significantly stronger than the T2K-only limit (orange-dashed curve), which appears due to complementary correlations of $\sin^2\eta$ and $\Delta m^2_{31}$. 

We observe from the right panels in \cref{fig:T2K+RENO-R12,fig:T2K+RENO-R3} a significant correlation between $\Delta m^2_{31}, \theta_{13}$ and $\sin^2\eta$. The averaging effects due to the oscillating neutrino masses can be partially compensated by adjusting $\Delta m^2_{31}$. However, comparing the upper-left and upper-right panels in \cref{fig:T2K+RENO-R12,fig:T2K+RENO-R3}, we see that in the acceptable range of $\sin^2\eta$, $\theta_{13}$ remains close to its vacuum value. The lower-left panels in the figures show the best-fit point of $\sin^2\theta_{23}$, which stays close to maximal mixing throughout the considered range. We remark that the noisy features visible for low values of $\sin^2\eta$ in both figures are not significant, as $\chi^2$ is very flat around maximal mixing.

\subsection{JUNO 59.1 days of data} \label{sec:JUNO}

Let us now discuss the constraints from the first 59.1 days of JUNO data \cite{JUNO:2025gmd}. Our analysis is based on a fit to the energy spectrum of the 2379 observed events following closely ref.~\cite{Esteban:2026phq}, where a detailed technical description of the analysis can be found. Given this significant number of events, we expect that any fast time oscillation will be sufficiently averaged out, justifying the treatment of time-averaging as discussed above. The exposure of 59.1~days corresponds to $\tau_\phi$ for $m_\phi \simeq 10^{-22}$~eV. Therefore, this first JUNO data sample is already sensitive to scalar masses consistent with the lower dark matter mass bound $m_\phi \gtrsim 10^{-21}$~eV. 

\begin{figure}
  \centering
  \includegraphics[width=0.65\textwidth]{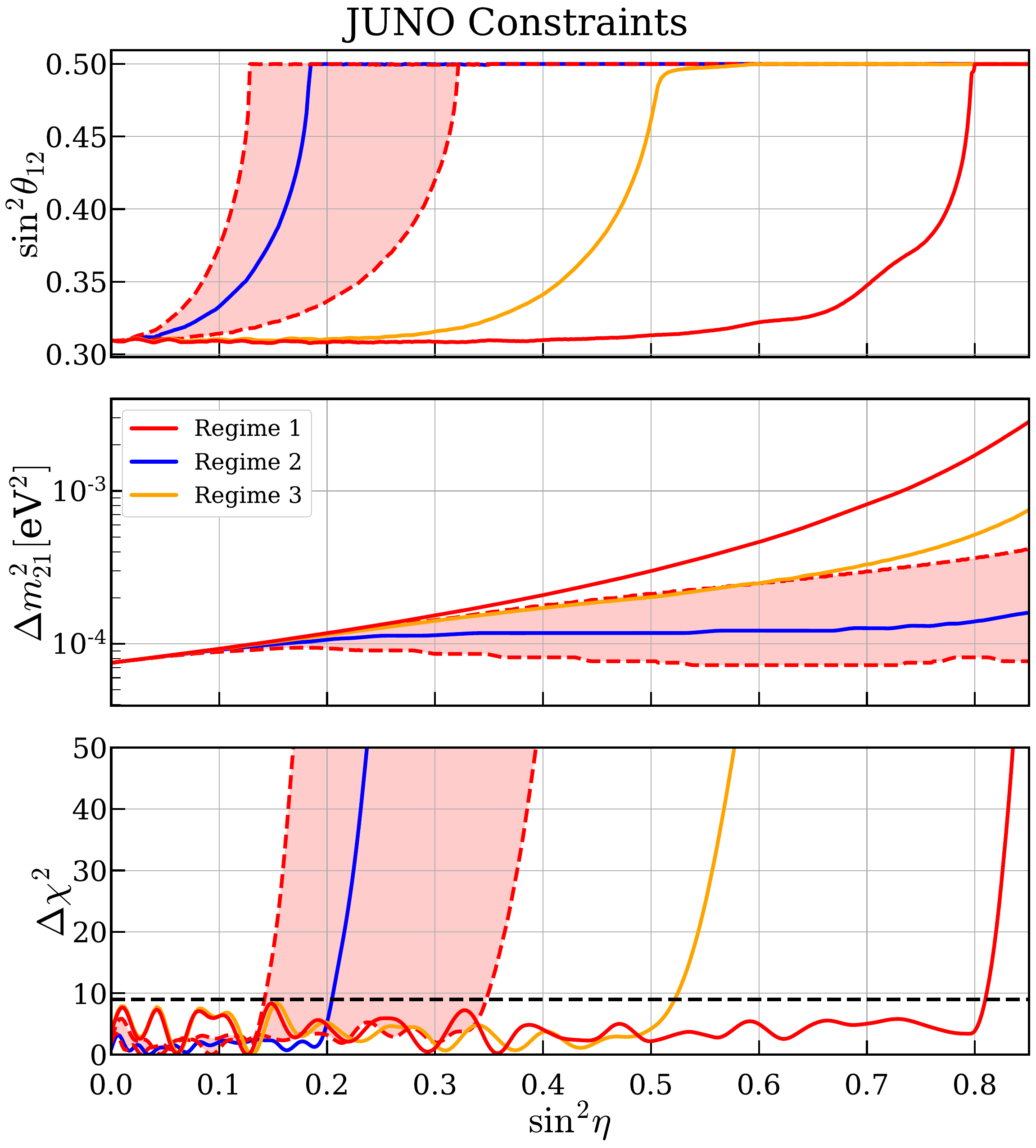}
  \mycaption{Constraints from 59.1 days of JUNO data for regimes 1, 2 and 3, shown in red, blue and orange, respectively. The bottom panel shows $\Delta\chi^2$ as a function of $\sin^2\eta$, the fraction of scalar-induced neutrino mass. The upper and middle panels show the best-fit values of the oscillation parameters $\theta_{12}$ and $\Delta m_{21}^2$ for each value of $\sin^2\eta$. The dashed curves indicate the $\pm \sigma(R)$ range for $R$ relevant for regime~1, whereas the solid red curve corresponds to $R=0.052$, which is the one-sided lower bound at 99.73\%~CL as obtained from the Rayleigh distribution.}
  \label{fig:JUNO}
\end{figure}

We show in \cref{fig:JUNO} the results of our analysis, focusing on the dominating oscillation parameters $\Delta m^2_{21}$ and $\theta_{12}$, as the current data set is not yet sensitive to the fast $\Delta m^2_{31}$-induced oscillations. We have verified that the results shown in \cref{fig:JUNO} do not significantly depend on the assumptions regarding $\Delta m^2_{31}$ and $\theta_{13}$. As a general trend, we observe in \cref{fig:JUNO} a strong correlation of the oscillation parameters with the scalar contribution to the neutrino mass. Up to $\Delta \chi^2$ fluctuations of a few units, the averaging effects can be compensated efficiently by increasing $\sin^2\theta_{12}$ and adjusting $\Delta m^2_{21}$: the damping of oscillations can be compensated by increasing their amplitude.
However, once maximal mixing $\sin^2\theta_{12}=0.5$ is reached, the oscillation amplitude $\sin^22\theta_{12}$ cannot be further enhanced. The averaging leads to a strong modification of the oscillation pattern (compare \cref{fig:probR12,fig:probR3}) and $\Delta \chi^2$ increases quickly with increasing $\sin^2\eta$. 

In regime 2, where both fast and slow oscillations are averaged out, we find that the scalar contribution to the neutrino mass is constrained to be less than about 20\% at $3\sigma$. However, the limit becomes relatively weak in regime~1, due to the unknown fluctuations of the $R$ parameter. Adopting the convention as above, we allow for a $3\sigma$ downward fluctuation to $R=0.052$ to set a $3\sigma$ bound. As can be seen in \cref{fig:probR12}, for such a low value of $R$ the dominating $\Delta m^2_{21}$-induced oscillation pattern in the survival probability remains relatively unaffected up to sizeable values of $\sin^2\eta$. Therefore, we find a relatively weak upper bound of $\sin^2\eta < 0.81$ at $3\sigma$ in regime~1, which, however, still strongly disfavours a 100\% dark origin of neutrino mass. In regime~3, for $2\times 10^{-11}$~eV~$\lesssim m_\phi \lesssim 3\times 10^{-8}$~eV, we obtain a bound of $\sin^2\eta < 0.52$ at $3\sigma$.

\subsection{Sensitivity of 6 years of JUNO data} \label{sec:JUNOsens}

The sensitivity of the first 59.1 days of JUNO data is limited to the dominating $\Delta m^2_{21}$-oscillation mode, as this data set is sensitive to the fast, $\Delta m^2_{31}$-driven oscillations only at the level of few $\Delta \chi^2$ units \cite{Esteban:2026phq}. This will change quickly with more exposure. Therefore, we investigate the sensitivity of the full planned six year exposure of JUNO. We rescale the signal and background event numbers of the 2025 data set  \cite{JUNO:2025gmd} with the appropriate factor and consider an Asimov data set (without statistical fluctuations) corresponding to pure vacuum neutrino masses (i.e., $\sin^2\eta=0$) and current best-fit oscillation parameters \cite{nufit-6.1}. Note that the longer exposure shifts the transition between regimes 1 and 2 to lower $m_\phi$ masses, whereas the transition between regimes 2 and 3 remains the same, as it is determined by the source-detector distance of the experiment. Similar as in the previous analysis, we keep $\theta_{13}$ fixed to its vacuum value. This is motivated by the T2K+RENO constraint, but we have checked that also marginalizing over it gives similar results. However, now $\Delta m^2_{31}$ plays a crucial role and for each value of $\sin^2\eta$ we search for the best-fit value, along with $\sin^2\theta_{12}$ and $\Delta m^2_{21}$.

\begin{figure}[t]
  \centering
  \includegraphics[width=0.9\textwidth]{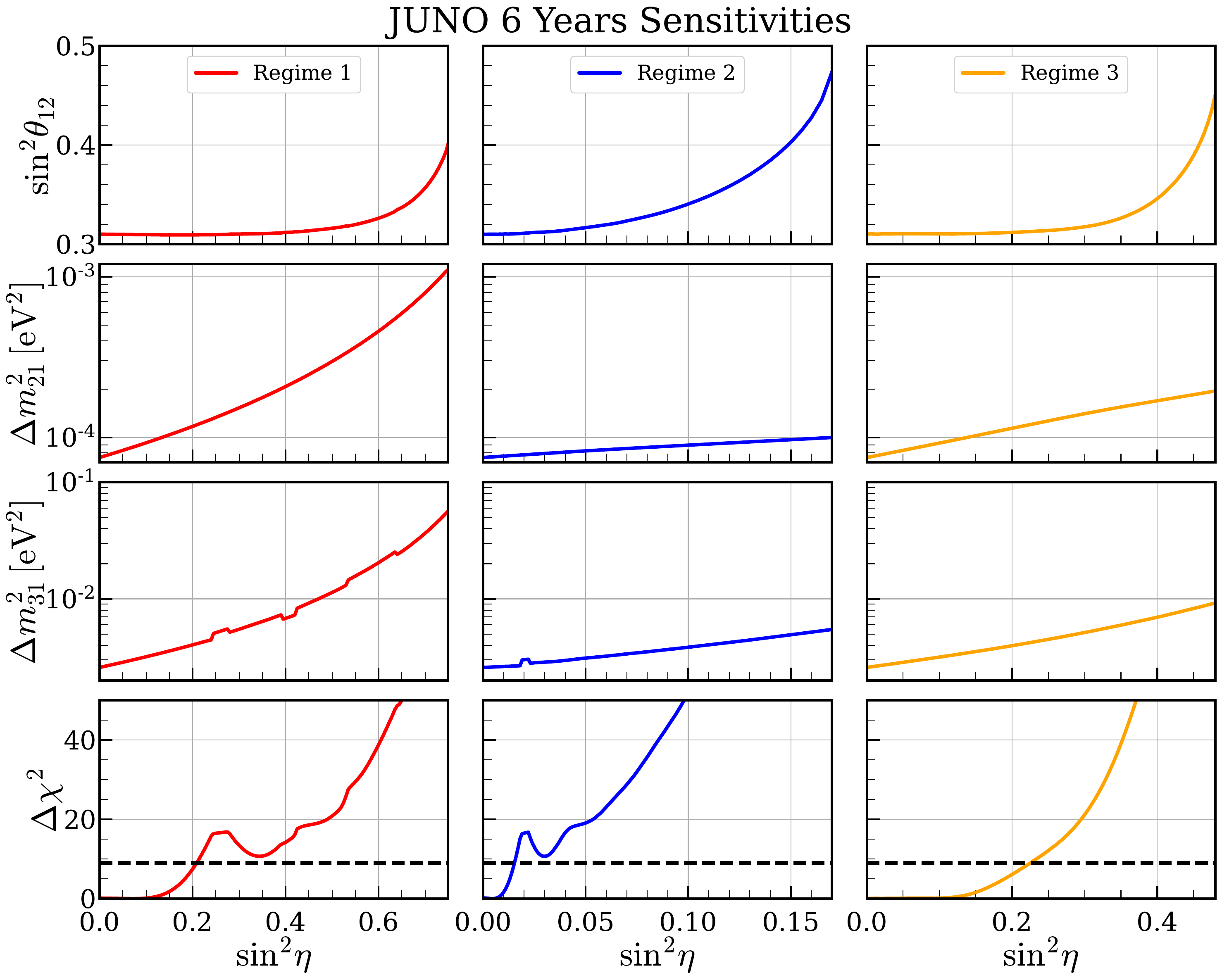}
  
  \mycaption{Sensitivity of 6 years of JUNO data for regimes 1, 2 and 3, shown in the left, middle, right columns, respectively. The bottom panels shows $\Delta\chi^2$ as a function of $\sin^2\eta$, the fraction of scalar-induced neutrino mass. The other panels show the best-fit values of the oscillation parameters $\theta_{12}$, $\Delta m_{21}^2$ and $\Delta m_{31}^2$ for each value of $\sin^2\eta$. $\theta_{13}$ is kept fixed at its vacuum value. Note the different scales on the horizontal axes.
  }
  \label{fig:JUNOsens}
\end{figure}

We show the results for the three regimes in \cref{fig:JUNOsens}. Considering first regime~2 (middle column of panels) we observe that $\sin^2\theta_{12}$ and $\Delta m^2_{21}$ behave very similar to the current data set shown in \cref{fig:JUNO}. However, now JUNO is sensitive with high significance also to the fast $\Delta m^2_{31}$-induced oscillations, which start getting damped already for smaller values of $\sin^2\eta$ (see \cref{fig:probR12}). Therefore, $\Delta\chi^2$ increases quickly, leading to the excellent sensitivity of $\sin^2\eta < 0.014$ at $3\sigma$. For regime~1 we observe a qualitatively similar behaviour of the $\Delta\chi^2$ curve, however, rescaled approximately as $\sin^2\eta  \to \sin^2\eta \times R/\langle R\rangle$ with $\langle R\rangle = 0.886$ and $R = 0.052$, see \cref{eq:bessel}. Correspondingly, the resulting sensitivity weakens significantly to $\sin^2\eta < 0.21$ at $3\sigma$, which, however, would still improve over the current best limit of 0.54 in regime~1. Similarly, also in regime~3 we find improved sensitivity due to the damping of $\Delta m^2_{31}$-induced oscillations, as illustrated in \cref{fig:probR3}, improving the limit on $\sin^2\eta$ from 0.52 for the 59.1 days exposure to 0.23. 

\section{Discussion and summary}\label{sec:discussion}

\begin{table}[t]
\centering
\small
  \begin{tabular}{c@{\quad}c@{\qquad}c@{\quad}c@{\qquad}c@{\quad}c}
    \hline\hline
     \multicolumn{2}{c}{T2K+RENO} & 
     \multicolumn{2}{c}{JUNO (59 d)} &
     \multicolumn{2}{c}{JUNO (6 yr)} \\
    \hline
    $m_\phi$ range [eV] & limit &
    $m_\phi$ range [eV] & limit &
    $m_\phi$ range [eV] & sens. \\
    \hline
    $2\cdot 10^{-21} - 9\cdot 10^{-17}$ & 0.54 &
    $2\cdot 10^{-21} - 2\cdot 10^{-15}$ & 0.81 &
    $2\cdot 10^{-21} - 7\cdot 10^{-17}$ & 0.21 \\
    $9\cdot 10^{-17} - 4\cdot 10^{-12}$ & 0.09 &
    $2\cdot 10^{-15} - 2\cdot 10^{-11}$ & 0.20& 
    $7\cdot 10^{-17} - 2\cdot 10^{-11}$ & 0.014\\    
    $4\cdot 10^{-12} - 9\cdot 10^{-10}$ & 0.23 &
    $2\cdot 10^{-11} - 3\cdot 10^{-8}$ & 0.52 &
    $2\cdot 10^{-11} - 3\cdot 10^{-8}$ & 0.23 \\    
    $9\cdot 10^{-10} - 6\cdot 10^{-9}$ & 0.37 & & & & \\
    \hline\hline
  \end{tabular}  
  \mycaption{$3\sigma$ upper bounds on the relative contribution of the dark matter scalar coupling to the neutrino mass, $\sin^2\eta$, in the different regimes of scalar masses $m_\phi$. The lowest limit on $m_\phi$ is the lower bound on the DM mass from \cite{Zimmermann:2024xvd}. We show the limits for the combined T2K+RENO analysis, for JUNO data from  59.1~days exposure, as well as the estimated sensitivity for a 6~year exposure in JUNO. The three ranges for JUNO correspond to regimes 1,2 and 3, see \cref{eq:regimes}. For T2K+RENO there is also the mixed case, where T2K is in regime~3 and RENO in regime~2, c.f.~\cref{fig:regimes}.}
  \label{tab:limits}
\end{table}

The main results of our analysis are summarized in \cref{tab:limits} and in \cref{fig:summary}. Over the whole range of scalar masses, from the lowest allowed value for a dark matter field, $2\times 10^{-21}$~eV, up to around $3\times 10^{-8}$~eV, we can exclude a pure scalar-induced neutrino mass at very high confidence level. This result is robust with respect to assumptions about coherence properties of the field and is based on the physically motivated assumption of a virialised dark matter state in the galaxy. Details of the dark matter distribution may only shift the $m_\phi$ values, where the transition between the different regimes occur. Our results would be somewhat modified only in the case of an extremely cold dark matter distribution, with negligible velocity spread, see discussion in \cref{sec:coherence}. This would not correspond to a virialised dark matter distribution but resemble more a Bose-Einstein condensate, dominated by a single momentum mode.

Our results apply to a real scalar or pseudo-scalar field, including the well-motivated axion-like particle case, which offers natural cosmological dark matter scenarios. Most of our results carry over also to a complex scalar field, which we discuss in some detail in appendix~\ref{app:complex}. The only exception is the case of a fully asymmetric scalar field, providing 100\% of neutrino masses, as indicated by the red bars in \cref{fig:summary}, which remains allowed in the region $2\times 10^{-21}$~eV~$\lesssim m_\phi\lesssim 10^{-16}$~eV. Here ``fully asymmetric'' means that only the particle or anti-particle component of the field is present in the background. We leave it as an open question, whether a plausible cosmological scenario exists, where such a configuration can be generated by a physical mechanism. 

In our analysis we have assumed that always one of the hierarchies of time scales as indicated in \cref{eq:regimes} holds. This means that all non-standard time variations due to the scalar field coupling are either very slow and can be treated as constant or fast enough that they are fully averaged. This implies that no anomalous variation of the event time structure is observable in the experiments. In the real data the time average has to be considered as a discrete sum over the time stamps of each neutrino event or a summation over time-binned data. Our approximation of perfect averaging holds for sufficiently large event numbers. Therefore, we focus on disappearance channels in our analysis, where event numbers are large and the time averaging approximation holds. 

For this reason, the upper bounds on $\sin^2\eta$ quoted in \cref{tab:limits} hold only sufficiently far away from the limits of the stated $m_\phi$ intervals. If some of the experimental and scalar time scales become comparable (i.e., close to the $m_\phi$-range boundaries in \cref{tab:limits} or the discrete jumps in \cref{fig:summary}) we expect a non-trivial time dependence in the data, an analysis of which would require detailed information on the time distribution of the data and is beyond the scope of this work. We note, however, that within our model assumptions we expect a smooth transition between the regimes 1, 2, 3 and one can expect that the limits are smoothly connected between the regimes. 

A simplifying assumption of our analysis has been that the coupling between scalar field and neutrinos is diagonal in the vacuum neutrino mass basis. This implies that lepton mixing angles are not affected by the scalar field.
In the right panel of \cref{fig:summary} we have translated our limits on $\sin^2\eta$ into constraints on the coupling constants $g_i$ between the scalar field and neutrinos by using \cref{eq:coupling_sqeta}, adopting the value $\rho_{\rm DM} = 0.4\,\rm GeV/cm^3$. For this conversion we assumed normal neutrino mass ordering and that the lightest neutrino mass $m_1$ is zero (and therefore also the coupling $g_1$). Then we have $m_i^0 = \sqrt{\Delta m^2_{i1}}$ ($i=2,3$) in \cref{eq:coupling_sqeta}. To derive the bound on $g_i$ we consider the strongest bound on $\sin^2\eta$ from the considered oscillation experiments. Then, we extract the best-fit values of $\Delta m^2_{i1}$ for this value of $\sin^2\eta$, which then is used in \cref{eq:coupling_sqeta} to convert the limit on $\sin^2\eta$ into a limit on $g_i$. 

If the lightest neutrino mass is non-zero, our constraint on the coupling constants $g_i$ becomes weaker. While the constraint on the relative scalar contribution $\sin^2\eta$ is independent of $m_1$, the couplings depend on the absolute mass scale via $g_i \propto \sqrt{\Delta m^2_{i1} + m_1^2} \sin^2\eta$ ($i=2,3$), see \cref{eq:coupling_sqeta}. Therefore, the limits in the left panel of \cref{fig:summary} as well as in \cref{tab:limits} are independent of the absolute neutrino mass scale, whereas the limits on $g_i$ shown in the right panel of \cref{fig:summary} are rescaled accordingly. 

Our analysis is based on a limited set of neutrino oscillation data, namely from the JUNO, T2K and RENO experiments. This choice is motivated by the following arguments: $(i)$ we are using disappearance data only, where large event numbers are available, such that time averaging is efficient, $(ii)$ to good approximation these data can be described by vacuum oscillations, which allows us to consistently assume that the neutrino--scalar coupling is diagonal in the same basis as the propagation basis of neutrino mass eigenstates and $(iii)$ while being minimal, this data set allows to constrain all oscillation parameters except the complex phase. With this choice we can derive conservative constraints within our phenomenological scenario. 

In general we find strong correlations of the dark contribution to the neutrino mass with the values of the oscillation parameters: for sizeable fractions of scalar-induced mass, both mixing angles and mass-squared differences deviate strongly from their vacuum values in order to compensate the averaging effects from the scalar oscillations. It remains to be shown whether these strong deviations remain viable in a more complete set of oscillation experiments, beyond the ones considered here.
Moreover, it can be expected that in more general scenarios for the scalar--neutrino interaction, additional effects may become relevant and lead to stronger constraints. For example, in general one expects also off-diagonal neutrino--scalar couplings in the mass basis, which can induce time-dependent mixing and additional flavour transitions (see e.g., \cite{Losada:2022uvr}). Similarly, in configurations where the MSW matter effect is relevant, additional non-trivial effects can be expected. We leave the investigation of this rich phenomenology for future work.

Finally, let us briefly comment on cosmological consequences. The authors of ref.~\cite{Aghaie:2026bqq} pointed out that the interaction \cref{eq:Lphi} leads to a loop-induced self-interaction potential for the scalar field (see also \cite{Dev:2022bae}), which is strongly constrained by cosmology, if $\phi$ provides the dark matter in the Universe \cite{Cembranos:2018ulm}. We indicate this constraint as the grey line in the right panel of \cref{fig:summary}. Moreover, if we take the dark neutrino mass generation mechanism at face value, we have $m_\nu \propto \sqrt{\rho_{DM}} \propto a^{-3/2}$, where $a$ is the cosmic scale factor. Hence, $m_\nu$ grows faster with redshift than the neutrino temperature, and the energy density in neutrinos scales as $\rho_\nu \propto m_\nu(a) a^{-3} \propto a^{-4.5}$, growing with redshift faster than radiation. Hence, neutrinos would dominate the energy density of the Universe roughly at a redshift $z \gtrsim 11$, significantly modifying  the expansion rate of the Universe during recombination and BBN. Therefore, one needs to invoke a mechanism to avoid the growth of the cosmological neutrino mass with redshift, see for instance \cite{Sen:2023uga,Sen:2024pgb,Huang:2022wmz}. 

\subsection*{Acknowledgement}

We thank Daniel Naredo-Tuero for useful discussions.
The work of S.S.C.\ is funded by the Deutsche Forschungsgemeinschaft (DFG, German Research Foundation), project number 510963981. 

\appendix

\section{Complex scalar field}\label{app:complex}

Let us now generalize \cref{eq:Lphi} to the case of a complex field $\phi$. Similar to the real case we again assume that the coupling $g_i$ is diagonal in the same basis as a possible bare neutrino mass term, but take it now to be complex in general. In analogy to \cref{eq:m_nu} we write
\begin{equation}
  \mu_i(\vec x,t) =
  m_i^0\left[c_\eta^2 + s_\eta^2 \mathcal{A}(\vec x,t)\right] \,,
\end{equation}
which now can be a complex mass parameter, with complex phases absorbed by $\mathcal{A}$. As a general parametrisation of the classical complex field we assume  
\begin{equation}\label{eq:Axt-compl}
  \mathcal{A}(\vec x,t) = \sum_{\vec k} \left[
  \alpha_{\vec k} \, a_{\vec k} \,\exp\left( -i\omega t +i \vec k \vec x + i\varphi_{\vec k} \right) +
  \beta_{\vec k} \, b_{\vec k} \,\exp\left( i \omega t  -i \vec k \vec x - i\varphi'_{\vec k} \right) 
  \right] \,,
\end{equation}
with 
\begin{equation}
\omega \approx m_\phi + \frac{\vec k^2}{2m_\phi}  \,,    
\end{equation}
and $\beta_{\vec k} , b_{\vec k},\varphi'_{\vec k}$ are defined in analogy to 
$\alpha_{\vec k} , a_{\vec k},\varphi_{\vec k}$ (see main text) but in general are independent of each other. If they are taken to be equal, we recover the real case. The complex masses will enter the neutrino evolution equation via a term \cite{Sen:2023uga}
\begin{equation}
    H \approx \frac{\mu\mu^\dagger}{2E_\nu} \,.
\end{equation}
Therefore, we need to replace \cref{eq:dmu2} by
\begin{align}
  \Delta \mu^2_{ij}(\vec x, t) & =
  \Delta m^2_{ij} \left|c_\eta^2 + s_\eta^2 \mathcal{A}(\vec x,t)\right|^2 \nonumber\\
  & = \Delta m^2_{ij} \left[
  c_\eta^4 + 2 c_\eta^2 s_\eta^2 \text{Re}[\mathcal{A}(\vec x,t)] + 
  s_\eta^4  |\mathcal{A}(\vec x,t)|^2\right]\,. \label{eq:dmu2-compl}
\end{align}

Let us now discuss, whether this expression can lead to qualitatively different phenomenology than the real case. First, if $s_\eta^2$ is small, the second term in the square-bracket of \cref{eq:dmu2-compl} will dominate over the third one proportional to $s_\eta^4$. The real part of $\mathcal{A}$ behaves qualitatively in the same way as the real version $A$ in \cref{eq:Axt}. In a similar way, it includes fast oscillations with typical frequencies set by $m_\phi$, as well as slow frequencies $\sim \vec k^2/m_\phi \sim 10^{-6} m_\phi$. Hence, for small $s_\eta^2$ we obtain a very similar behaviour for the real and complex cases and we expect that all numerical results apply as well for the complex case.

Let us consider now the case of $s_\eta^2 \simeq 1$, i.e., the coupling to the dark matter scalar gives the dominant contribution to the neutrino mass and the last term in \cref{eq:dmu2-compl} dominates. For the same reason as discussed in the main text, we can absorb the $\vec k \vec x$ terms into the phases $\varphi_{\vec k}$ and $\varphi'_{\vec k}$. Keeping only the time-dependent terms we have
\begin{align}
    |\mathcal{A}(t)|^2 &= \sum_{\vec k, \vec q} \left[
    \tilde a_{\vec k} \tilde a_{\vec q} \, e^{-i \frac{\vec k^2-\vec q^2}{2m_\phi}t}
    e^{i(\varphi_{\vec k} - \varphi_{\vec q})} +
    \tilde b_{\vec k} \tilde b_{\vec q} \, e^{i \frac{\vec k^2-\vec q^2}{2m_\phi}t}
    e^{-i(\varphi'_{\vec k} - \varphi'_{\vec q})}
    \right] \nonumber\\
    & +  \sum_{\vec k, \vec q} \left[ 
    \tilde a_{\vec k}\tilde b_{\vec q}  
    \, e^{-2im_\phi t}e^{-i \frac{\vec k^2+\vec q^2}{2m_\phi}t}
    e^{i(\varphi_{\vec k} + \varphi'_{\vec q})} + \text{c.c.} 
    \right] \label{eq:Asq-compl}
\end{align}
with 
$\tilde a_{\vec k} \equiv \alpha_{\vec k}  a_{\vec k}$ and
$\tilde b_{\vec k} \equiv \beta_{\vec k}  b_{\vec k}$. Again we observe that in general this has the same time structure as $A^2(t)$ from \cref{eq:At} and our numerical results will apply also in the complex case. 

There is, however, one exception leading to a qualitatively different result, namely if either $a_{\vec k}$ or $b_{\vec k}$ vanishes. In this case, \cref{eq:Asq-compl} will contain no fast oscillating terms. Physically, this corresponds to a fully asymmetric scalar background, consisting only of particles or antiparticles. In this case we find
\begin{equation}
        |\mathcal{A}(t)|^2 = \left| \sum_{\vec k} 
        \tilde a_{\vec k} \, e^{-i \frac{\vec k^2 t}{2m_\phi} + i\varphi_{\vec k}}
        \right|^2 = R^2(t)  \qquad\text{(asymmetric background)}\,, 
\end{equation}
with $R^2(t)$ defined in \cref{eq:R_gamma_beta}. Hence, this leads to the same behaviour as in regime~3 as described in \cref{eq:R3a,eq:R3}. Note that for $s_\eta^2=1$, \cref{eq:R3} becomes
\begin{align}\label{eq:R3-compl}
   \left\langle \cos\frac{\Delta \mu^2_{ij}L}{2E_\nu} \right\rangle_{T_{\rm exp}} =
   \frac{1}{1 + \Delta_{ij}^2 /4}  \,,
\end{align}
with no oscillatory dependence on $L/E_\nu$. 
This result applies, as long as $T_{\rm exp} \gg \tau_v$, i.e., the ``slow'' oscillations due to the $\vec k^2/(2m_\phi)$ terms are still fast on the time scale of the experiment. Hence, we can apply the results obtained for the real scalar in regime (R3) with $s^2_\eta=1$ to the case of the asymmetric complex scalar in regimes (R2) and (R3). The only case where our constraints do not apply, is the asymmetric complex scalar in regime (R1). In this case, $R(t)$ is constant at the time scale of the experiment, and oscillations will appear as in vacuum with an effective mass-squared difference set by $\Delta \mu^2_{ij} = \Delta m^2_{ij} R^2$, with $R^2$ being random, but constant on the time scale of the experiment.

\section{Details of the T2K simulation}
\label{app:t2k}

Below, we briefly describe the experimental specifications and the numerical analysis method of the long-baseline experiment T2K. T2K is an off-axis accelerator-based neutrino experiment with a 295~km baseline between the source and the far detector. The (anti)muon neutrino beams are produced at J-PARC facility in Tokai using a 30~GeV proton beam and detected in the Super-Kamiokande Water Cherenkov detector with a fiducial volume of 22.5~kton. In our numerical analysis, we closely follow the experimental configurations described in~\cite{T2K:2023smv,T2K:2023mcm,T2K:2025wet}. The data sets we used are based on $1.97\times 10^{21}$ ($1.63\times 10^{21}$) protons on target (P.O.T.) in the neutrino (antineutrino) mode \cite{T2K:2023smv,T2K:2023mcm}.

As discussed in the main text, in this work we consider only the muon neutrino and antineutrino disappearance channels, which are primarily sensitive to the atmospheric oscillation parameters $\theta_{23}$  and $\Delta m_{31}^2$. The dominant background in these disappearance channels arises from neutral current interactions. To account for the systematic uncertainties we assign an uncorrelated 5\% (2\%) signal normalization (energy calibration) error, a 5\% (2\%) background normalization (energy calibration) error for the $\nu$-mode. For the $\bar{\nu}$-mode, where the statistics are comparatively lower and systematic uncertainties are larger, we apply a 10\% (10\%) signal normalization (energy calibration) error, a 10\% (10\%) background normalization (energy calibration) error. The neutrino and antineutrino fluxes reported in~\cite{T2K:2023smv} are used with a proper normalization factor based on the reported protons on target. For estimating the cross sections of different neutrino interactions we follow the same reference as for the fluxes. To reproduce the energy-dependent event spectra at the far detector as close as possible, we employ properly optimized smearing matrices for the energy resolution and post-smearing matrices for the efficiency of the detector. 

To fit the data within the theoretical framework, we use the GLoBES~\cite{Huber:2004ka,Huber:2007ji} package with the necessary modifications of the oscillation probability engine, glb files as well as the chi-squared functions. In the analysis the standard oscillation parameters $\theta_{23}$ and $\Delta m^2_{31}$ are freely marginalized over while keeping the leptonic CP-phase fixed at $\delta_{\rm CP} = 0^{\circ}$. We have verified that varying $\delta_{\rm CP}$ to other representative values has a negligible impact on our final results because of the limited sensitivity of the disappearance channels to this phase. The treatment of $\theta_{13}$ and the solar oscillation parameters are discussed in \cref{sec:T2K-RENO}.

In order to estimate the sensitivity to both the standard and new physics parameters, we adopt a frequentist statistical approach and minimize the following $\chi^2$-function:
\begin{equation}\label{eq:chi2}
  \chi^2(\lambda, \lambda^\prime) =  \min\limits_{\xi_x}\left[2\sum_i
    \left(N_i(\lambda,\,\lambda^\prime, \, \xi_x) -N_i^{\rm obs} - N_i^{\rm obs}\, \ln \frac{N_i(\lambda,\,\lambda^\prime, \,\xi_x)}{N_i^{\rm obs}}\right) + \sum_{x = 1}^4 \frac{\xi_x^2}{\sigma_{\xi_x}^2} \right]
    \,.
\end{equation}
Here $i$ labels the reconstructed energy bin, $\lambda$ denotes the model parameters held fixed in the analysis, and $\lambda^\prime$ are the parameters over which we marginalize. $N_i(\lambda,\,\lambda^\prime,\,\xi_x)$ is the number of predicted events in the $i$-th bin computed including both the signal and background contributions along with the related various systematics uncertainties described above. The systematics are parametrized by pull parameters $\xi_x$ with the associated errors $\sigma_{\xi_x}$.
$N_i^{\rm obs}$ corresponds to the ``observed'' number of events in the $i$-$th$ bin as reported by the T2K collaboration.
The final $\chi^2$ for each channel is obtained by minimizing over the oscillation parameters $\lambda^\prime$ and then we define the test statistic $\Delta\chi^2$ projected onto the fixed parameter(s) $\lambda$ as
\begin{equation}\label{eq:Delchi2}
 \Delta\chi^2(\lambda) =   \min\limits_{\lambda^\prime}\left[ \chi^2(\lambda, \lambda^\prime)\right] - \min\limits_{\lambda} \left[ \chi^2(\lambda, \lambda^\prime) \right].
\end{equation}

As a necessary check of our analysis we have verified, that we can reproduce the standard oscillation analysis of 
\cite{T2K:2023smv,T2K:2023mcm} to good accuracy, and our confidence regions in the oscillation parameters are in good agreement with the published ones. 

Throughout the fitting procedure we have assumed a vacuum oscillation probability for the disappearance channels, motivated by our assumptions on the coupling of neutrinos to the scalar field. This is a well-justified approximation for the T2K baseline, where the matter effect is small, see also \cite{Denton:2024thm}. Moreover, since we marginalize freely over the two most important parameters of interest for this channel, namely $\theta_{23}$ and $\Delta m_{31}^2$, any residual matter-induced modifications to the oscillation probability are effectively absorbed into the fitted values of these parameters.

\paragraph{Estimate of effective exposure time.} The averaging of the fast oscillations due to the scalar field coupling in the experimental data should be considered as a discrete sum over the time stamps of the neutrino events. In the case of an accelerator experiment, and T2K in particular, the time structure of the events is controlled by the beam power, the beam-on versus beam-off periods as well as different event rates for neutrinos and anti-neutrinos. In order to estimate the effective exposure time for T2K we proceed as follows. Based on Fig.~1 of \cite{T2K:2023smv}, we construct a probability distribution of the events in time, $f(t)$, using beam power (in POT) as a function of time, weighted with the neutrino versus anti-neutrino event rate per POT, and removing the beam-off periods. From this distribution we can calculate an effective time spread of the data by 
\begin{equation}
    \delta t = \sqrt{\langle t^2 \rangle - \langle t \rangle^2} \,,
    \qquad
    \langle t^n \rangle  = \int dt \, t^n \, f(t) \,.
\end{equation}
We take $T_{\rm exp} = 2\delta t$ as an estimate for the effective exposure time for T2K, yielding $T_{\rm exp} \approx 1.6$~yr as quoted in \cref{tab:times}. 

\bibliographystyle{JHEP_improved}
\bibliography{./refs}

\end{document}